\documentclass[aps,nofootinbib,preprint,superscriptaddress,floatfix]{revtex4-1}

\usepackage[USenglish]{babel}
\usepackage{epsfig}
\usepackage{amssymb}
\usepackage{amsmath}
\usepackage{bm}
\usepackage{hyperref}
\usepackage{enumerate}
\usepackage{graphicx,picture,calc}

\begin{document}

\title{Competing electronic instabilities of extended Hubbard models on the honeycomb lattice: A functional Renormalization Group calculation with high wavevector resolution}

\author{D. S\'anchez de la Pe\~na}
\affiliation{Institute for Theoretical Solid State Physics, RWTH Aachen University, D-52074 Aachen, Germany}
\author{J. Lichtenstein}
\affiliation{Institute for Theoretical Solid State Physics, RWTH Aachen University, D-52074 Aachen, Germany}
\author{C. Honerkamp}
\affiliation{Institute for Theoretical Solid State Physics, RWTH Aachen University, D-52074 Aachen, Germany}
\affiliation{JARA-FIT, J\"ulich Aachen Research Alliance - Fundamentals of Future Information Technology}

\date{\today}

\begin{abstract}
We investigate the quantum many-body instabilities for electrons on the honeycomb lattice at half-filling with extended interactions, motivated by a description of graphene and related materials. We employ a recently developed fermionic functional Renormalization Group scheme which allows for highly resolved calculations of wavevector dependences in the low-energy effective interactions. We encounter the expected anti-ferromagnetic spin density wave for a dominant on-site repulsion between electrons, and charge order with different modulations for dominant pure $n$-th nearest neighbor repulsive interactions. Novel instabilities towards incommensurate charge density waves take place when non-local density interactions among several bond distances are included simultaneously. Moreover, for more realistic Coulomb potentials in graphene including enough non-local terms there is a suppression of charge order due to competition effects between the different charge ordering tendencies, and if the on-site term fails to dominate, the semi-metallic state is rendered stable. The possibility of a topological Mott insulator being the favored tendency for dominating second nearest neighbor interactions is not realized in our results with high momentum resolution. 
\end{abstract}


\maketitle

\section{INTRODUCTION}

Since the experimental realization of graphene, its unique properties have taken the spotlight due to a wide range of promising applications. It also constitutes a theoretical playground for unusual many-body phenomena\cite{Geim2007,CastroNeto2009}. We analyze the possible groundstates of extended Hubbard models on the two-dimensional honeycomb lattice at charge neutrality, focusing on implications for single-layer graphene. Considerable effort has been put through a diversity of methods to address this matter\cite{Sorella1992,Herbut2006,Vafek2007,Uchoa2007,Honerkamp2008,Herbut2009,Herbut2009a,Meng2010,Sorella2012,Ulybishev2013,Classen2014,Hohenadler2014,Golor2015,Tang2015}.

In this work, we employ the newly developed \textit{Truncated Unity functional Renormalization Group} scheme (TUfRG)\cite{Lichtenstein2016}, building on prior channel-decomposed fRG\cite{Husemann2009} and SMfRG\cite{Wang2012} schemes. The cornerstone of fRG methods for interacting fermions is their unbiasedness in comparison with mean-field treatments or single-channel calculations like the \emph{random phase approximation}. The fRG treats all ordering tendencies on equal footing, and directly provides a description of the effective low-energy degrees of freedom without the need for prior assumptions about the dominating low-energy correlations. Previous results within the widely used \textit{Fermi surface patching} fRG scheme\cite{Zanchi1998,Honerkamp2001,Halboth2000} were computationally limited in their resolution of wavevector dependences of the arising effective interactions. The TUfRG scheme has enabled us to increase the wavevector resolution in a highly scalable\cite{Lichtenstein2016,Proceedings} and numerically efficient way. The main motivation behind a finer momentum resolution of effective interactions in graphene is twofold. On the one hand, there is evidence about the influence of wavevector resolution on qualitative predictions of fRG results\cite{Scherer2015,Kiesel2012,Wang2012,Kiesel2013,Wang2013}, particularly for the realization of exotic topological groundstates in the honeycomb lattice\cite{Raghu2008,Scherer2015}. A higher momentum-resolved calculation could shed light on the inconclusive fate of such topologically non-trivial phases. On the other hand, an attempt to handle the unscreened nature of Coulomb interactions in graphene at half-filling requires an approach able to resolve the momentum dependence of bare interactions going beyond the first few nearest neighbor repulsion terms.

In sections \ref{sec:mod} and \ref{sec:method} we introduce the model considered and the TUfRG scheme, respectively. In section \ref{sec:res1} of this paper, we present a tentative phase diagram for electrons in the honeycomb lattice subjected to on-site, first and second nearest neighbor repulsive interactions. Not only do we find a strong suppression of the topological quantum Hall state compared to previous fRG results, but also incommensurate charge orderings arise which had not yet been observed. In section \ref{sec:res2}, after including a third nearest neighbor coupling term and using realistic ab-initio interaction parameters for graphene\cite{Wehling2011}, we find that the semi-metallic state might well be stabilized by competing charge ordering tendencies.

\section{MODEL}\label{sec:mod}

The relevant physics in our system of interest can be captured by extended repulsive Hubbard models for interacting electrons in a honeycomb lattice. The non-interacting part is described by a tight-binding Hamiltonian with nearest neighbor hoppings and at half-filling (i.e. $\mu=0$)

\begin{equation}
 H_0 = -t \sum_{<i,j>,s} \left(c^{\dagger}_{i,s,A}\,c_{j,s,B} + \mathrm{h.c.} \right)\,,
\end{equation}
where operators $c^{(\dagger)}_{i,s,o}$ annihilate (create) an electron at lattice site $i$ with spin $s$ in orbital $o$. By orbital degrees of freedom we are referring to sublattice indices, not to be confused with the different orbitals within a single atom. The hopping amplitude $t$ sets the energy unit relative to which the interaction strengths and energy scales will be expressed. The inter-atomic distance between nearest neighbors is normalized to unity. This kinetic part can be diagonalized in momentum space to reveal two energy bands featuring two inequivalent Dirac cones at the $\mathbf{K,K'}$ points in the Brillouin zone corners. At half-filling, the single-particle density of states vanishes at the Fermi level due to the presence of linear band crossing points in the low-energy dispersion. The vanishing density of states translates to a semi-metallic behavior which is stable against spontaneous symmetry breaking tendencies induced by interactions, at least up to some finite critical interaction strengths. It also implies that interaction processes away from the Fermi level play a more relevant role in comparison with cases where the density of states stays finite or even diverges at the Fermi level. Thus a fine Brillouin zone discretization is required, as evidenced in Ref. \onlinecite{Scherer2015}.

The interacting part of the Hamiltonian up to second nearest neighbor reads

\begin{equation}
 H_{\mathrm{int}}= U \sum_{i,o} n_{i,\uparrow,o} \, n_{i,\downarrow,o} + V_1 \! \sum_{\substack{<i,j> \\ s,s'}} n_{i,s,A} \, n_{j,s',B} + V_2 \!\!\! \sum_{\substack{<<i,j>> \\ s,s',o}} \!\! n_{i,s,o} \, n_{j,s',o}
\end{equation}
where $n_{i,s,o}=c^{\dagger}_{i,s,o}\,c_{i,s,o}$ are local density operators, $<i,j>$ and $<<i,j>>$ represent first and second nearest neighbor pairs respectively. 

%

Since the RG flow will be calculated in the band picture, where the kinetic part is diagonal, $H_{\mathrm{int}}$ also has to be transformed from orbital to band degrees of freedom. That unitary transformation produces some extra momentum structure for the bare interactions, the so-called \textit{orbital makeup}, which also has to be properly sampled in momentum space (see appendix).

\section{METHOD}\label{sec:method}

\begin{figure}[htb]
  \centering
  \begin{tabular}{@{}cc@{}}
   \hspace*{.8cm} \includegraphics[width=.21\textwidth]{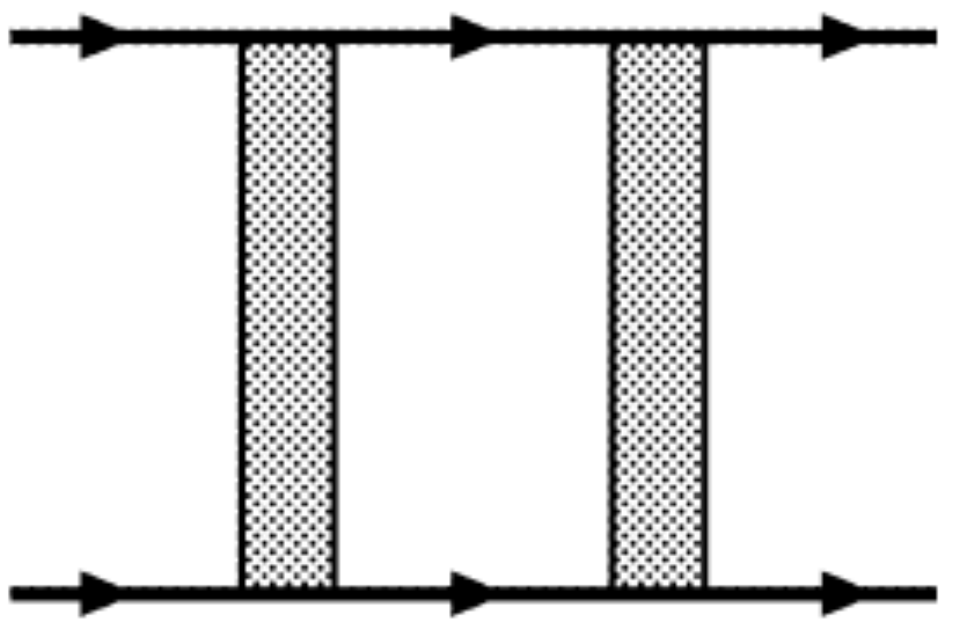} \hspace*{.2cm} &
    \includegraphics[width=.21\textwidth]{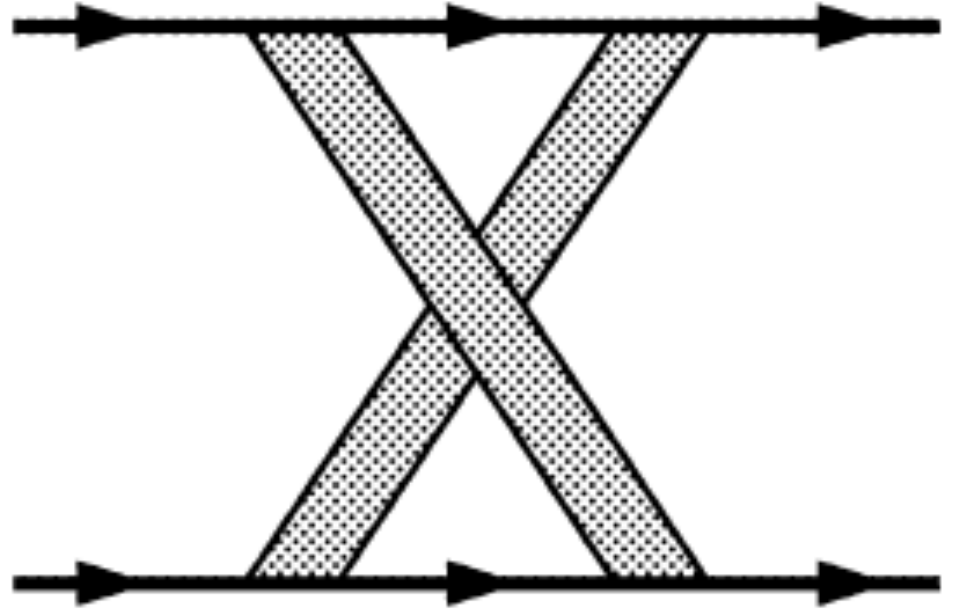}   \\[3ex]
    \multicolumn{2}{c}{\includegraphics[width=.55\textwidth]{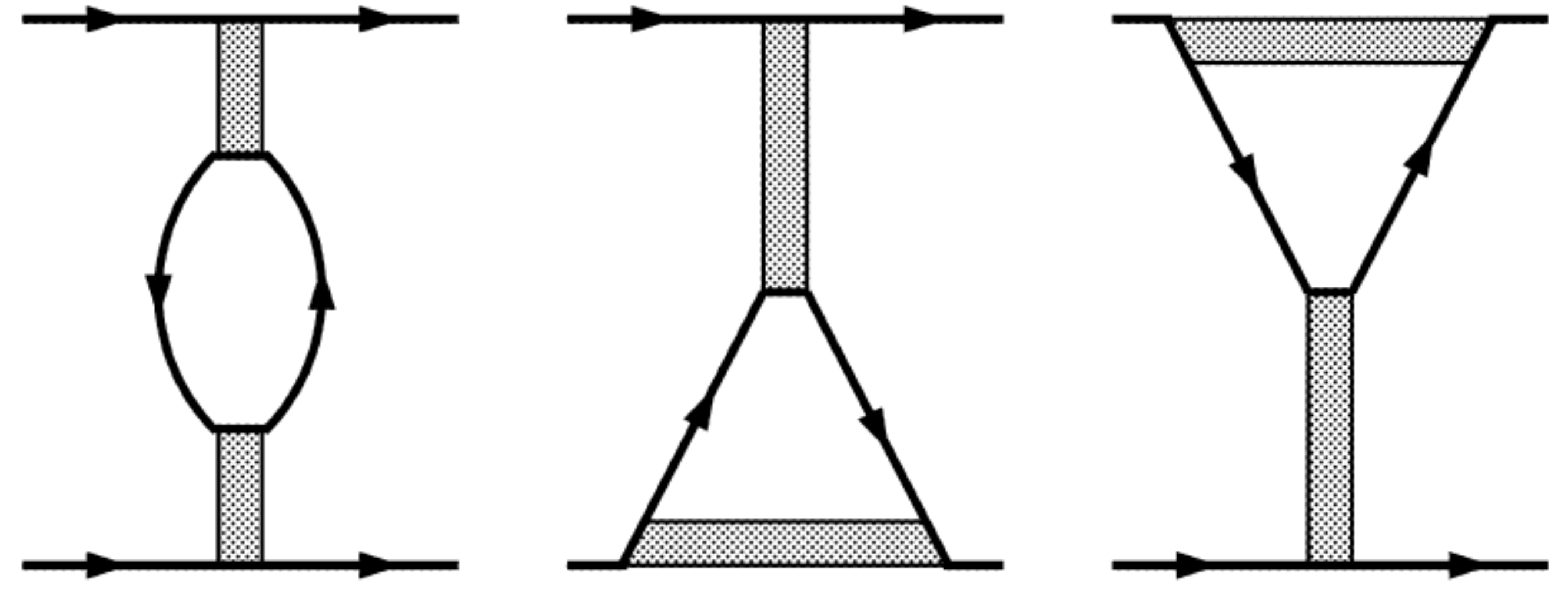}}
  \end{tabular}
  \caption{Diagrammatic representation of the right-hand side of Eq.(\ref{eq:flow}), with particle-particle (top left), crossed particle-hole (top right) and direct particle-hole (bottom) diagrams.}
  \label{fig:diag}
\end{figure}

We employ a functional renormalization group method within the one-loop, one-particle-irreducible (1PI) formalism for fermionic systems\cite{Wetterich1993,Metzner2012,Thomale2013} to perform a weak-coupling instability analysis. The choice of regulator follows the so-called $\Omega$ scheme\cite{Husemann2009}, i.e. infrared divergences are regularized by a soft frequency cutoff. The free Green's function $G_0(\omega,\mathbf{k},b)$ for Matsubara frequency $\omega$, wavevector $\mathbf{k}$ and band index $b$ gets modified by including a regulator $C_\Omega(\omega,\mathbf{k})$ that depends on an auxiliary energy scale $\Omega$ which discriminates high and low-energy modes

\begin{align*}
 G_0(\omega,\mathbf{k},b) \rightarrow G_{0,\Omega} (\omega,\mathbf{k},b) & = C_\Omega(\omega) G_0(\omega,\mathbf{k},b)\,, \\
 C_\Omega(\omega) & = \frac{\omega^2}{\omega^2+\Omega^2}\,.
\end{align*}
In contrast to commonly used momentum-shell cutoff schemes, this does not artificially suppress small-momentum particle-hole fluctuations. The inclusion of a scale dependent regulator makes the generating functional for 1PI vertex functions scale dependent as well, $\Gamma \rightarrow \Gamma_\Omega$. The functional flow equation obtained by differentiating $\Gamma_\Omega$ respect to $\Omega$ allows to interpolate between the well-known microscopic bare action and the sought-after effective action at low energies. The functional flow equation boils down to a coupled infinite hierarchy of flow equations for the vertex functions, as revealed by a power expansion of $\Gamma_\Omega$ in the fermionic fields. For computational feasibility, the hierarchy is truncated after the two-particle vertex function and the flow of the self-energy is neglected. Such approximations are justifiable in weak-coupling regimes\cite{Salmhofer2001}. The two-particle vertex function for a U(1) and SU(2) invariant theory reads

\begin{align*}
 \Gamma^{(4)}_\Omega [\bar{\psi},\psi] & = \frac{1}{2} \int \! d \xi_1 \dots d \xi_4 \, V_\Omega^{b_1 \dots b_4}(k_1,k_2,k_3)\, \delta(k_1+k_2-k_3-k_4) \\
 & \qquad \times \sum_{s,s'} \bar{\psi}_s (\xi_4) \, \bar{\psi}_{s'} (\xi_3) \, \psi_{s'} (\xi_2) \, \psi_s (\xi_1) \, ,
\end{align*}
where $k_i = (\omega_i,\mathbf{k}_i)$ and $\xi_i = (\omega_i,\mathbf{k}_i,b_i)$ are multi-index quantum numbers containing a Matsubara frequency $\omega_i$, wavevector $\mathbf{k}_i$ and band index $b_i$, and $\int d \xi_i$ is shorthand notation for $\int \frac{d \mathbf{k}_i}{A_{B\!Z}}\frac{1}{\beta}\sum_{\omega_i} \sum_{b_i}$ with Brillouin zone area $A_{B\!Z}$ and inverse temperature $\beta$. The vertex function can then be parametrized by a spin independent coupling function $V_\Omega^{b_1 \dots b_4}(k_1,k_2,k_3)$ which depends on three frequencies, three wavevectors and four band indices. From now on, dependences on the regularization scale will not be explicitly written and dot notation will be used for scale derivatives. The flow equation for the coupling function consists of three channels

\begin{equation} \label{eq:flow}
 \dot{V}^{\{b_i\}} (k_1,k_2,k_3) = \mathcal{T}_\mathrm{pp}^{\{b_i\}} (k_1,k_2,k_3)
 +  \mathcal{T}^{\mathrm{cr}, \, \{b_i\}}_\mathrm{ph}  (k_1,k_2,k_3) + \mathcal{T}^{\mathrm{d}, \, \{b_i\}}_\mathrm{ph} (k_1,k_2,k_3)
\end{equation}
corresponding to particle-particle, crossed and direct particle-hole contributions respectively (Fig. \ref{fig:diag}). They constitute the one-loop diagrams of second order in the interaction which are allowed by the symmetries of the system

\begin{align} \label{eq:diag}
 \mathcal{T}_\mathrm{pp}^{\{b_i\}} = - \int \! d p & \, \left[ \partial_\Omega \,G(p,b) \, G(k_1+k_2-p,b') \right] \notag \\ 
 & \times  V^{b_1 b_2 bb'} (k_1,k_2,p) \,V^{bb'b_3 b_4}(k_1+k_2-p,p,k_3)\,, \notag \\[1.0ex] 
 \mathcal{T}^{\mathrm{cr}, \, \{b_i\}}_\mathrm{ph} = - \int \! d p & \, \left[ \partial_\Omega \,G(p,b) \, G(p+k_3-k_1,b') \right]  \notag \\
 & \times V^{b_1 b'bb_4} (k_1,p+k_3-k_1,k_3) \, V^{bb_3b_2 b'}(p,k_2,p+k_3-k_1)\,,  \\[1.0ex] 
 \mathcal{T}^{\mathrm{d}, \, \{b_i\}}_\mathrm{ph} = - \int \! d p & \,\, \left[ \partial_\Omega \,G(p,b) \, G(p+k_2-k_3,b') \right]  \notag \\  
& \times \left[ -2 V^{b_1 bb_3 b'} (k_1,p+k_2-k_3,p) \, V^{b'b_2 bb_4}(p,k_2,k_3) \right. \notag \\
&  +  V^{b_1 b' b b_3} (k_1,p+k_2-k_3,k_1+k_2-k_3) \, V^{bb_2 b' b_4}(p,k_2,k_3) \notag  \\  
& \left. +  V^{b_1 b' bb_4} (k_1,p+k_2-k_3,p) \, V^{b b_3 b_2 b'}(p,k_2,p+k_2-k_3) \right]\,,  \notag
\end{align}

where $\int d p=\int \frac{d \mathbf{p}}{A_{B\!Z}}\frac{1}{\beta}\sum_{\omega} \sum_{b b'}$. Since we are interested in static ground state properties, the flow equation is solved for a frequency independent interaction, with external frequencies set to zero. The remaining dependences on external momenta are dealt with by means of the Truncated Unity scheme\cite{Lichtenstein2016}, a highly scalable formalism based on channel-decomposed flows\cite{Husemann2009,Husemann2012,Eberlein2010,Wang2012,Eberlein2013,Maier2013,Xiang2013,Eberlein2014}. Since the singular contributions to the flow are produced for specific transfer momenta between external legs in the one-loop diagrams of Fig. \ref{fig:diag}, the coupling function $V$ can be well parametrized by a decomposition into single-channel coupling functions 

\begin{align} \notag
 V^{\{b_i\}} \left(\mathbf{k}_1,\mathbf{k}_2,\mathbf{k}_3 \right)   & = V^{(0),\, \{b_i\}}_{\mathbf{k}_1,\mathbf{k}_2,\mathbf{k}_3}  - \Phi^{\mathrm{SC},\,\{b_i\}}_{\mathbf{k}_1+\mathbf{k}_2,\frac{\mathbf{k}_1-\mathbf{k}_2}{2},\frac{\mathbf{k}_4-\mathbf{k}_3}{2}}  + \Phi^{\mathrm{C},\, \{b_i\}}_{\mathbf{k}_1-\mathbf{k}_3,\frac{\mathbf{k}_1+\mathbf{k}_3}{2},\frac{\mathbf{k}_2+\mathbf{k}_4}{2}} \\  &
 \quad + \Phi^{\mathrm{D}, \{b_i\}}_{\mathbf{k}_3-\mathbf{k}_2,\frac{\mathbf{k}_1+\mathbf{k}_4}{2},\frac{\mathbf{k}_2+\mathbf{k}_3}{2}}\,, \, 
\end{align}
each picking up a corresponding dependence on each of the three respective transfer momenta appearing in the diagrams. $V^{(0)}$ is the initial bare interaction and stays constant, whereas the $\Phi$s are generated during the flow according to

\begin{align}
 \dot{\Phi}^{\mathrm{SC},\,\{b_i\}}_{\mathbf{k}_1+\mathbf{k}_2,\frac{\mathbf{k}_1-\mathbf{k}_2}{2},\frac{\mathbf{k}_4-\mathbf{k}_3}{2}} & =  - \mathcal{T}_\mathrm{pp}^{\{b_i\}}  \left(\mathbf{k}_1,\mathbf{k}_2,\mathbf{k}_3 \right)\,,   \notag \\
 \dot{\Phi}^{\mathrm{C},\,\{b_i\}}_{\mathbf{k}_1-\mathbf{k}_3,\frac{\mathbf{k}_1+\mathbf{k}_3}{2},\frac{\mathbf{k}_2+\mathbf{k}_4}{2}} & =  \mathcal{T}_\mathrm{ph}^{\mathrm{cr}, \, \{b_i\}}  \left(\mathbf{k}_1,\mathbf{k}_2,\mathbf{k}_3 \right)\,,  \\
 \dot{\Phi}^{\mathrm{D},\,\{b_i\}}_{\mathbf{k}_3-\mathbf{k}_2,\frac{\mathbf{k}_1+\mathbf{k}_4}{2},\frac{\mathbf{k}_2+\mathbf{k}_3}{2}} & =   \mathcal{T}_\mathrm{ph}^{\mathrm{d}, \, \{b_i\}}  \left(\mathbf{k}_1,\mathbf{k}_2,\mathbf{k}_3 \right)\,. \notag
\end{align}
The strong dependence on transfer momenta (first argument) is then kept as if effectively carried by an auxiliary exchange boson, while the two weak dependences on other momenta are expanded onto a suitable form factor basis

\begin{align}\label{eq:expansion}
 \Phi^{\mathrm{SC},\,\{b_i\}}_{\mathbf{l},\mathbf{k},\mathbf{k'}} & = \sum_{m,n} f_m(\mathbf{k}) \, f^*_n(\mathbf{k'}) \, P^{\{b_i\}}_{m,n} (\mathbf{l})\,, \notag \\
 \Phi^{\mathrm{C},\,\{b_i\}}_{\mathbf{l},\mathbf{k},\mathbf{k'}} & = \sum_{m,n} f_m(\mathbf{k}) \, f^*_n(\mathbf{k'}) \, C^{\{b_i\}}_{m,n} (\mathbf{l})\,,  \\
 \Phi^{\mathrm{D},\,\{b_i\}}_{\mathbf{l},\mathbf{k},\mathbf{k'}} & = \sum_{m,n} f_m(\mathbf{k}) \, f^*_n(\mathbf{k'}) \, D^{\{b_i\}}_{m,n} (\mathbf{l})\,.  \notag
\end{align}
This reparametrization is formally exact so far, but in practice the infinite basis of form factors has to be truncated. A right choice of form factor basis minimizes the truncation error. The coupling function can then be conveniently described by three objects, each having just one momentum dependence. Upon discretization of the strong momentum dependences, the numerical effort scales linearly respect to the number of sampling points, in contrast with a cubic scaling if discretizing $V^{b_1 \dots b_4}(k_1,k_2,k_3)$ directly. However, decomposing the couplings that appear in the right-hand side of Eqs.~(\ref{eq:flow})-(\ref{eq:diag}) in an attempt to rewrite the flow in terms of the three bosonic propagators $P,C,D$ generates intricate diagrams which are challenging to compute. The complication is due to internal bosonic lines appearing in the fermionic loops and thus having to integrate over the bosonic propagators of these internal lines, which eventually become sharply peaked during the flow. Moreover, the presence of the bosonic propagators in the integrands hinders the parallel scalability of a computer code implementation of these integrals. Such difficulties are avoided in the TUfRG by inserting partitions of unity of the form factor basis at the internal lines of the one-loop diagrams\cite{Lichtenstein2016}, isolating projections of the coupling function onto the form of the three channels ($V^\mathrm{P,C,D}$). In practice the partition of unity is truncated, so the TUfRG entails an additional approximation compared to other channel decomposed schemes. The insertion of a partition of unity allows for a simpler diagrammatic structure by pulling out the internal bosonic lines, turning Eqs.~(\ref{eq:flow})-(\ref{eq:diag}) into the TUfRG flow equations

\begin{align}
 \dot P_{m,n}^{\{b_i\}}(\mathbf{l}) & = \sum_{m',n'} \sum_{b,b'} V^{\mathrm{P},\,b_{1}b_{2}bb'}_{m,m'}\!(\mathbf{l})\,\dot \chi^{\mathrm{pp},\,bb'}_{m',n'}\!(\mathbf{l})\,V^{\mathrm{P},\,bb'b_{3}b_{4}}_{n',n}(\mathbf{l})\,, \notag \\[1.0ex]
 \dot C_{m,n}^{\{b_i\}}(\mathbf{l}) & = \sum_{m',n'} \sum_{b,b'} -V^{\mathrm{C},\,b_{1}b'bb_{4}}_{m,m'}(\mathbf{l})\,\dot \chi^{\mathrm{ph},\,bb'}_{m',n'}\!(\mathbf{l})\,V^{\mathrm{C},\,bb_{3}b_{2}b'}_{n',n}\!(\mathbf{l})\,, \notag   \\[1.0ex]
  \dot D_{m,n}^{\{b_i\}}(\mathbf{l}) & = \sum_{m',n'} \sum_{b,b'} \Big( 2V^{\mathrm{D},\,b_{1}bb_{3}b'}_{m,m'}\!(\mathbf{l})\,\dot \chi^{\mathrm{ph},\,bb'}_{m',n'}\!(\mathbf{l})\,V^{\mathrm{D},\,b'b_{2}bb_{4}}_{n',n}(\mathbf{l})\,-  \\
 &\quad -V^{\mathrm{C},\,b_{1}b'bb_{3}}_{m,m'}(\mathbf{l})\,\dot \chi^{\mathrm{ph},\,bb'}_{m',n'}\!(\mathbf{l})\,V^{\mathrm{D},\,bb_{2}b'b_{4}}_{n',n}(\mathbf{l})-V^{\mathrm{D},\,b_{1}bb_{3}b'}_{m,m'}\!(\mathbf{l})\,\dot \chi^{\mathrm{ph},\,bb'}_{m',n'}\!(\mathbf{l})\,V^{\mathrm{C},\,b_{2}b'bb_{4}}_{n',n}(\mathbf{l})\Big)\,, \notag
\end{align}
where 
\begin{align} \label{eq:chis}
 \chi^{\mathrm{pp},\,bb'}_{m,n} (\mathbf{l}) &= \int \! dp \, G \left(\omega_p\,, \frac{\mathbf{l}}{2} + \mathbf{p},b \right) \, G \left(-\omega_p\,, \frac{\mathbf{l}}{2} - \mathbf{p},b' \right) \, f^*_m (\mathbf{p}) \, f_n (\mathbf{p}) \,, \notag \\[0.5ex]
 \chi^{\mathrm{ph},\,bb'}_{m,n} (\mathbf{l}) &= \int \! dp \, G \left(\omega_p\,, \mathbf{p} + \frac{\mathbf{l}}{2},b \right) \, G \left(\omega_p\,, \mathbf{p} - \frac{\mathbf{l}}{2},b' \right) \, f^*_m (\mathbf{p}) \, f_n (\mathbf{p}) \,, 
\end{align}
and
\begin{align} \label{eq:projs}
  V^{\mathrm{P},\,\{b_i\}}_{m,n}\left(\mathbf{l} \right) & = \hat{P} \left[ V^{\{b_i\}} \right]_{m,n} (\mathbf{l})\,,  \notag \\
  V^{\mathrm{C},\,\{b_i\}}_{m,n}\left(\mathbf{l} \right) & = \hat{C} \left[ V^{\{b_i\}} \right]_{m,n} (\mathbf{l})\,,   \\
  V^{\mathrm{D},\,\{b_i\}}_{m,n}\left(\mathbf{l} \right) & = \hat{D} \left[ V^{\{b_i\}} \right]_{m,n} (\mathbf{l})\,. \notag
\end{align}
In the loop integrals $\chi^{\mathrm{pp}}, \,\chi^{\mathrm{ph}}$ the bosonic propagators have been replaced by slowly varying form-factors, with the additional advantage that its $m,n,\mathbf{l}$ components can all be calculated independently from each other. The price to pay is having to calculate the so-called \textit{inter-channel projections} above in \ref{eq:projs}, which are less computationally demanding and are also parallelized in a sacalable way. Projection operators $\hat{P},\hat{C},\hat{D}$ are defined in the appendix, and act as an inverse to the expansions (\ref{eq:expansion}). For a more detailed derivation of the scheme and its computational advantage see Ref. \onlinecite{Lichtenstein2016}. 

At the start of the flow, the values for the projected $V$s are the corresponding projections of the bare coupling $V^{(0)}$. The bosonic propagators start at value zero, and they pick up the renormalized corrections to the bare coupling during the flow as modes are integrated out by successively reducing the RG scale $\Omega$. If the system under consideration has a non-Fermi liquid as ground state after including electron-electron interactions, the normal metallic phase may become unstable towards a symmetry broken state. Such instabilities manifest themselves as divergences of some effective coupling components when $\Omega$ is lowered below a given value $\Omega_*$. If the flow of the self-energy were to be considered, these divergences or \textit{flows to strong coupling} would not take place since self-energy corrections to the electronic dispersion (e.g. a gap opening in the low-energy spectrum) would keep vertex-functions regular. In a level-2 truncation of the hierarchy without self-energy corrections, the flow has to be stopped at $\Omega_*$ due to the breakdown of the approximations used. Nevertheless, a weak-coupling instability analysis allows us to extract some relevant physical information about the possible ground states. The stopping point $\Omega_*$ provides an estimate for the critical scale $\Omega_C$, and the kind of coupling components which diverge signals the type of symmetry broken phase the system might enter. 

Technically speaking, the flow starts with the actual microscopic bare interaction for $\Omega \rightarrow \infty$ and the full effective interaction is recovered for $\Omega \rightarrow 0$. We typically start the flow at a scale two orders of magnitude bigger than that of the single-particle bandwidth, and stop the flow when the leading coupling component exceeds the order of magnitude of such bandwidth. The precise choice of a stopping point has no relevant effect on $\Omega_*$ as long as the initial $\Omega$ is big enough, since the couplings diverge strongly as the instability is approached.

A divergence in the $P$ channel indicates a pairing instability, divergences in the $C$ channel imply a tendency towards magnetic ordering, and charge order tendencies are encoded in the so-called $K$ channel
\begin{align}
 \Phi^{\mathrm{K},\,\{b_i\}}_{\mathbf{k}_3-\mathbf{k}_2,\frac{\mathbf{k}_1+\mathbf{k}_4}{2},\frac{\mathbf{k}_2+\mathbf{k}_3}{2}} = & - 2  \Phi^{\mathrm{D},\,\{b_i\}}_{\mathbf{k}_3-\mathbf{k}_2,\frac{\mathbf{k}_1+\mathbf{k}_4}{2},\frac{\mathbf{k}_2+\mathbf{k}_3}{2}} +\Phi^{\mathrm{C},\,\{b_i\}}_{\mathbf{k}_3-\mathbf{k}_2,\frac{\mathbf{k}_1+\mathbf{k}_4}{2},\frac{\mathbf{k}_2+\mathbf{k}_3}{2}}\,, \notag \\[1.0ex] 
  \Phi^{\mathrm{K},\,\{b_i\}}_{\mathbf{l},\mathbf{k},\mathbf{k'}} = & \sum_{m,n} f_m(\mathbf{k}) \, f^*_n(\mathbf{k'}) \, K^{\{b_i\}}_{m,n} (\mathbf{l})\,.  
\end{align}
The ordering vector is pinpointed by the momentum $\mathbf{l}$ at which the respective exchange propagator becomes sharply peaked.

\begin{figure}[htp]
\centering

\sbox0{\includegraphics[width=0.65\textwidth]{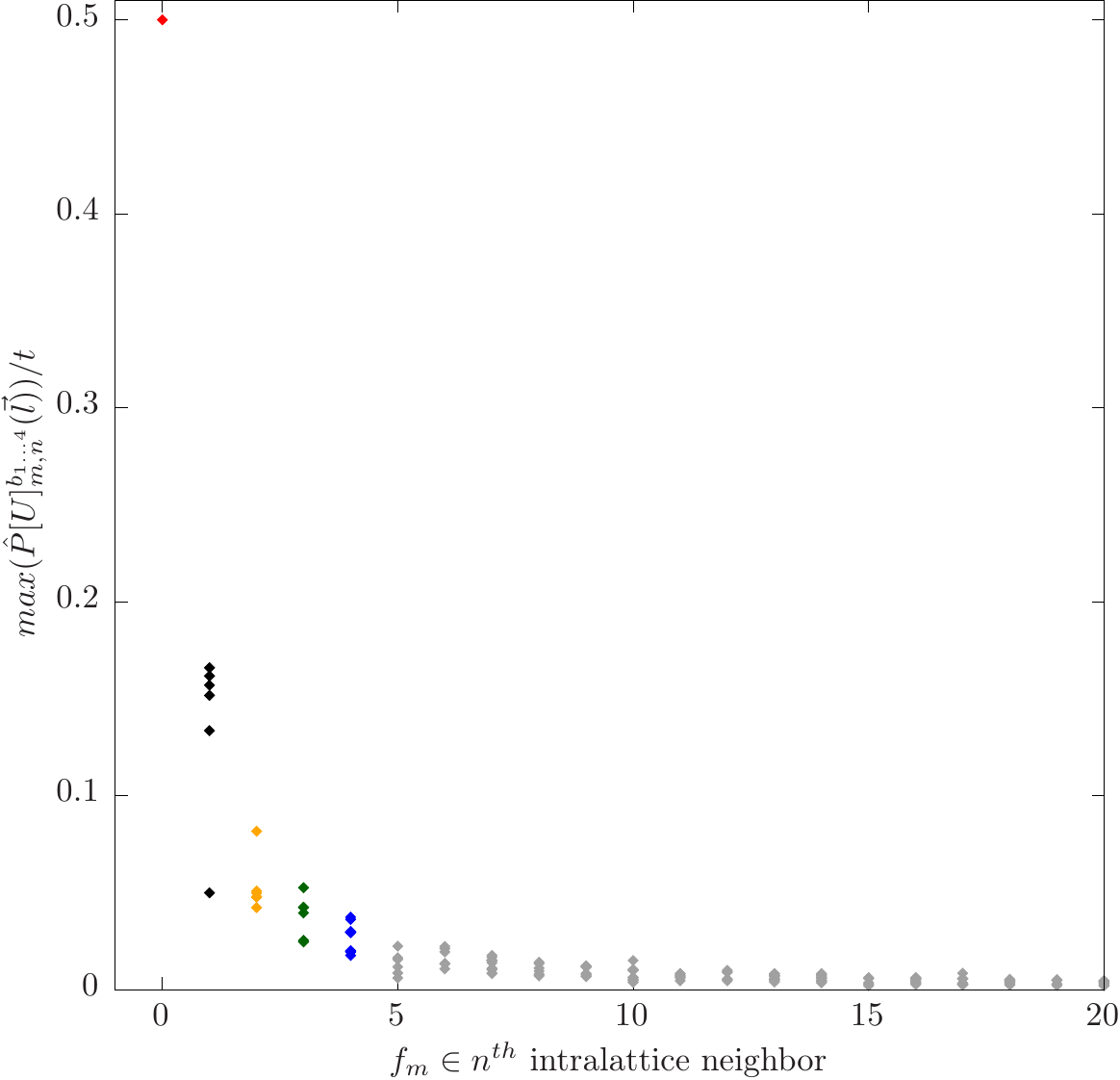}}
\sbox2{\includegraphics[width=0.35\textwidth]{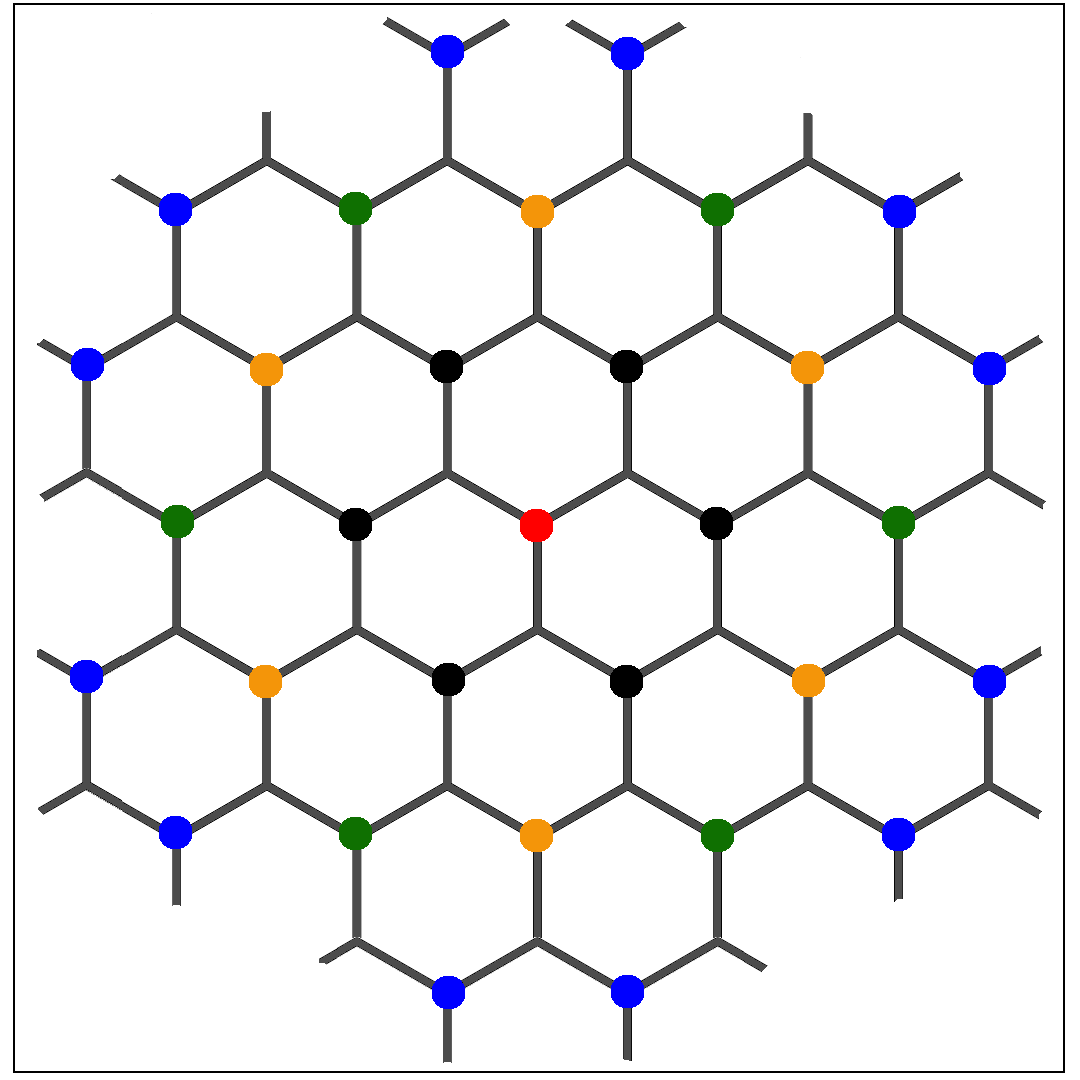}}

\begin{picture}(\wd0,\ht0)
\put(0,0){\usebox0}
\put(\wd0 - \wd2 - 0.5cm,\ht0 - \ht2 - 0.5cm){\usebox{2}}
\end{picture}
\caption{Maximal component (out of all choices for $f_n,\mathbf{l},\{b_i\}$) of the pairing channel projection of an on-site bare interaction $U=1\,t$ for $f_m$'s belonging to different nearest intra-lattice neighbor shells in real space.}
\label{fig:init_proj}
\end{figure}
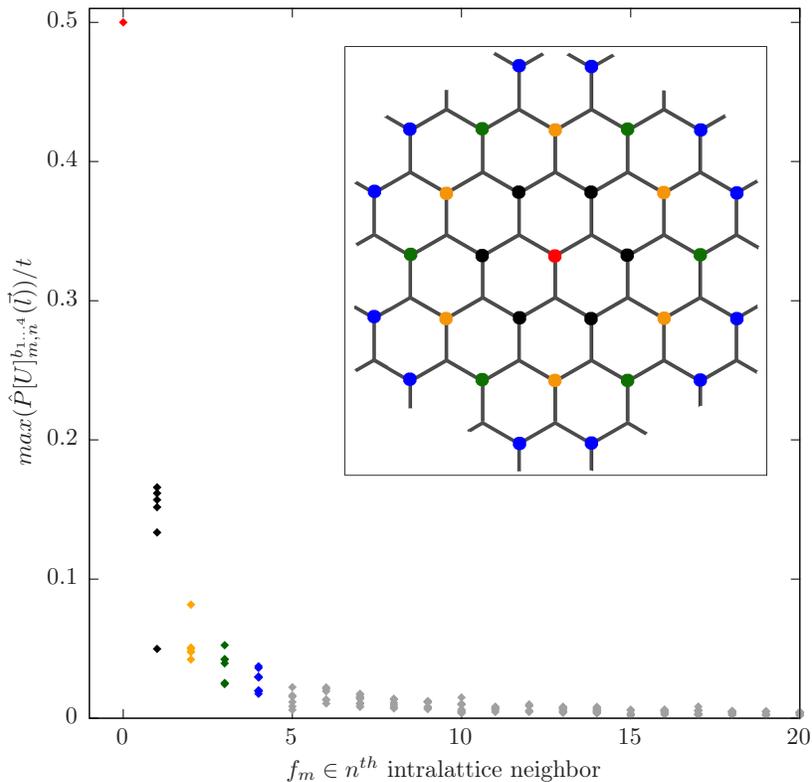

The form factor basis used here consists of sixfold symmetric lattice harmonics, defined directly in band picture. In real space they are localized $\delta$-peaks at fixed bond positions of the triangular Bravais lattice, grouped in shells by distance (see inset of Fig. \ref{fig:init_proj}) and transforming according to the irreducible representations of the $C_{6v}$ point group. One can choose the form factors to be real valued in momentum space. Since dependences on non-transfer momenta at weak-coupling exhibit no sharp features in momentum space, they are well described by slowly varying form factors i.e. form factors corresponding to short bond distances in real space. Truncation errors become unimportant once enough shells of $n$-th nearest neighbor bonds have been included in the form factor basis. With this choice of basis, which follows the irreducible representations of the lattice point group, the corresponding form factor indices at which divergences occur reveal the symmetry of the order parameter to be induced at the phase transition. 

Describing the flow in band picture is more economical than in orbital picture. In orbital picture the Green's functions are not diagonal, and form factors are more intricate since they must carry orbital indices. Point group symmetry operations do not mix band indices but may map sublattices onto one another, affecting the orbital indices. On the other hand, working in the band picture entails some challenges for projecting the bare interactions. Having a bare interaction that acts at a given bond distance in the lattice, the projections in orbital picture amount to just calculating overlaps of $\delta$-functions in position space. In band picture however, the orbital to band transformation elements add some extra structure. As a consequence, there is a non-zero overlap with form factors of any bond distance. As one would expect, the weight sits mainly at the form factor components that match the bond distance at which the interaction is acting in orbital language. Fortunately, they decay quickly enough if the distances differ (see Fig. \ref{fig:init_proj}), and truncating the basis after the first few nearest neighbors should capture short-ranged bare interactions rather accurately. For further details see the appendix.

As already mentioned, the fRG method constitutes an unbiased tool for studying the interplay and competition between different ordering tendencies towards a symmetry broken state. There is no need for an educated guess about the low-energy states, neither a need to single out some specific kind of correlations to be included in the renormalization procedure. Solving the flow equation amounts to an infinite order unbiased summation of all possible combinations of the arising particle-particle and particle-hole diagrams, so that all possible correlations are treated on equal footing. The following effective low-energy Hamiltonians presented in the results are no ansatz, they arise spontaneously in the flow depending on the choice for the bare interactions. 

\section{RESULTS}

\begin{figure}
 \centering
   \begin{tabular}{@{}c|c@{}}
    \includegraphics[width=.47\textwidth]{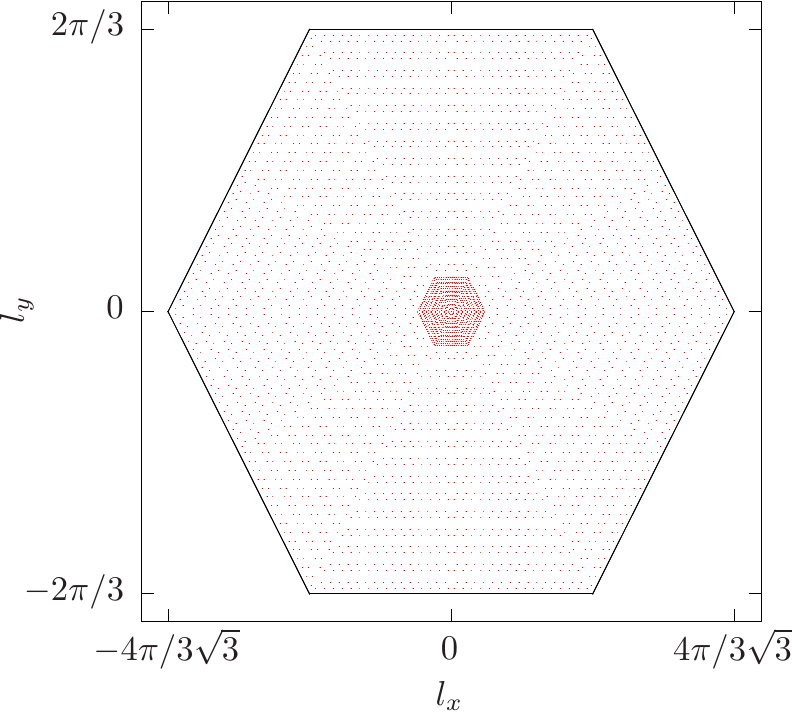}\hspace*{.1cm} &
    \hspace*{.1cm}\includegraphics[width=.47\textwidth]{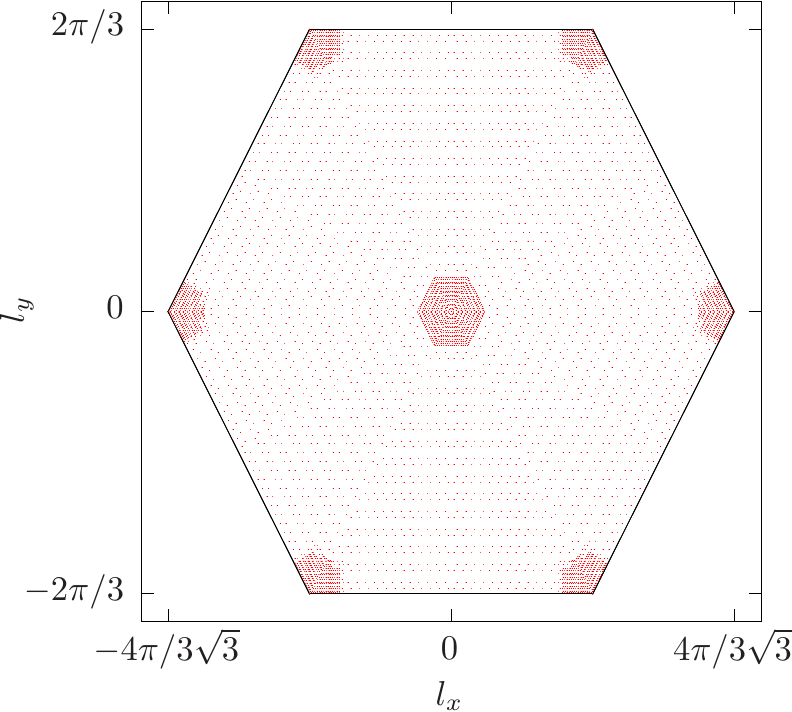}   \\
  \end{tabular}
  \caption{Example discretization for the dependence on transfer momenta $\mathbf{l}$, denser where the ordering vectors are expected. \textit{Left:} Mesh of $N_k = 3217$ points for momentum transfers in the particle-particle channel. \textit{Right:} Mesh of $N_k = 3661$ points for momentum transfers in particle-hole channels, specifically the one used for a pure $V_2$ bare interaction. For pure onsite and pure $V_1$ interactions, the mesh used for momentum transfers in particle-hole channels is the same as that used in the particle-particle channel.}
  \label{fig:mesh}
 
\end{figure}

In this work we present the results of weak-coupling instability analyses for the honeycomb lattice at half-filling and zero temperature within a TUfRG approach. The TUfRG flow equations are solved numerically by a discretization of wavevector dependences in the Brillouin zone into $N_k$ regions, reducing the integro-differential flow equation to a coupled system of $N_b^4 \times N_{ff}^2 \times N_k$ non-linear ordinary differential equations, where $N_b$ is the number of bands and $N_{ff}$ is the number of form factor functions. The ODE system is then solved using a fifth order Adams-Bashforth method. The transfer momenta are discretized into meshes of typically over 3200 points for the particle-particle channel and over 3600 points for particle-hole channels, as shown in Fig. \ref{fig:mesh}. The form factor basis is truncated after the second shell of nearest intra-lattice neighbors (fifth nearest real neighbor). In convergence tests we have included form factors up to the fourth shell, and meshes of up to 5000 points for momentum transfers.

The different tendencies towards symmetry broken ground states are characterized and a tentative phase diagram is obtained. We also provide estimates for the critical scales at which such transitions may occur. Possible deviations due to the approximations involved in our scheme are also discussed.

\subsection{Emerging instabilities}\label{sec:res0}
\noindent \textit{$\bullet$ Anti-ferromagnetic spin density wave} (SDW) \textit{instability}

\begin{figure}
\centering

 \includegraphics[width=0.6\textwidth]{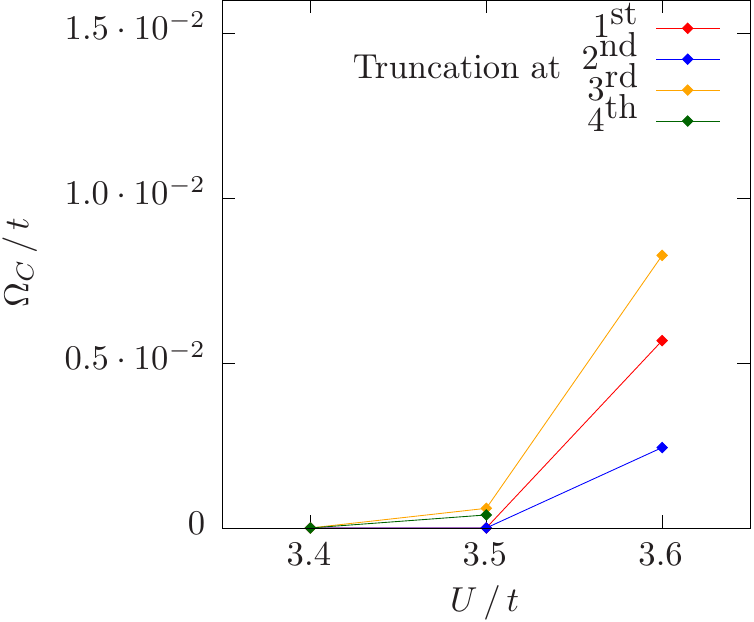}
 
\caption{Critical coupling strength for a pure on-site interaction with different truncations of the form-factor basis. Calculations including up to the fourth shell of form-factors are costly and thus have not been computed for $U=3.6t$. The reduction in $U_C$ for truncations including further neighbors can be understood as an effect of contributions coming from higher lattice harmonics.}
\label{fig:Uc}
 \end{figure}

This tendency is driven by an on-site bare interaction exceeding a critical value $U_C \approx 3.5\,t$ (see Fig. \ref{fig:Uc}). It manifests itself in the flow as a divergence in the magnetic propagator at zero momentum transfer and s-wave form factor components. The low-energy effective Hamiltonian obtained reads

\begin{equation}
H_{\text{SDW}} = -\frac{1}{\mathcal{N}} \sum_{o,o'}V_{o,o'}\epsilon_{o}\epsilon_{o'}\mathbf{S}^{o}\cdot \mathbf{S}^{o'}
\end{equation}
with $\mathbf{S}^{o}=\frac{1}{2} \sum_{\mathbf{k},s,s'} \mathbf{\sigma}_{s,s'} c^{\dagger}_{\mathbf{k},s,o} c_{\mathbf{k},s',o}$, $V_{o,o'}>0$ and $\epsilon_{o} = +1$ for $o \in \{\text{A}\}$, $\epsilon_{o} = -1$ for $o \in \{\text{B}\}$. The interaction becomes infinitely ranged, and is attractive for intra-sublattice scatterings and repulsive for inter-sublattice scatterings. The system adopts anti-ferromagnetic order as opposite net spin moments are induced on the different sublattices. The spin quantization axis is not fixed. This transition opens a gap in the electronic spectrum.

As a consequence of numerics, the precise choice for the unitary transformation from orbital to band degrees of freedom affects the resulting value for the critical coupling strength. The value shown above is fortuitously near the exact numerical results (of about $3.8\,t$) although fRG calculations are expected to underestimate critical coupling strengths by a wider margin, due to the neglect of bosonic collective fluctuations. As discussed in the appendix, with a different choice of orbital makeup a value of $U_C \approx 2.7\,t$ is obtained, in much better agreement with the most recent and finely discretized Fermi-surface patching results available\cite{Yanick2016}. This matter has only a quantitative effect on results and does not play a role in the qualitative discussion that follows.

\noindent \textit{$\bullet$ Charge density wave} (CDW,$\mathrm{CDW}_3$,iCDW's) \textit{instabilities}

We find several types of charge order emerging in the honeycomb lattice model. A nearest neighbor bare interaction over a critical value drives a conventional charge density wave (CDW), signaled by diverging couplings in the charge channel with zero momentum transfer and s-wave form factor components. The low-energy effective Hamiltonian is

\begin{equation}
H_{\text{CDW}} = -\frac{1}{\mathcal{N}} \sum_{o,o'}V_{o,o'}\epsilon_{o}\epsilon_{o'} N^{o}N^{o'}
\end{equation}
with $N^{o}=\sum_{\mathbf{k},s} c^{\dagger}_{\mathbf{k},s,o} c_{\mathbf{k},s,o}$. The orbital sign structure is the same as in the previous instability, which here translates to an infinitely ranged attraction for sites on the same sublattice and repulsion between different sublattice sites. Consequently, the system develops a higher charge occupancy in one of the sublattices. In this phase the energy spectrum becomes gapped as well.

Another charge ordering is found with an enlarged unit cell, named as three-sublattice charge density wave ($\mathrm{CDW}_3$) due to the splitting of each sublattice intro three with redistributed charge densities. It is driven by a supercritical second nearest neighbor bare interaction $V_2$, and shows up as a divergence in the charge channel with momentum transfer $\mathbf{Q} = \mathbf{K} - \mathbf{K}'$ and s-wave form factor components. The low-energy effective Hamiltonian becomes

\begin{equation}
H_{\text{CDW}_3} = -\frac{1}{\mathcal{N}} \sum_{o,o'}V_{o,o'}\epsilon_{o}\epsilon_{o'} (N^{o}_{\mathbf{Q}}N^{o'}_{\mathbf{-Q}}+N^{o}_{\mathbf{-Q}}N^{o'}_{\mathbf{Q}})
\end{equation}
with $N^{o}_{\mathbf{Q}}=\sum_{\mathbf{k},s} c^{\dagger}_{\mathbf{k+Q},s,o} c_{\mathbf{k},s,o}$ and the same orbital sign structure once again. In this case there is a modulated charge occupancy of the form $\,\,\sim \cos(\mathbf{Q}\cdot \mathbf{R} + \alpha)$ for lattice site $\mathbf{R}$, and depending on a phase factor $\alpha$ which controls the relative charge distribution between the three emergent sublattices. A more detailed description of the mean-field order parameter and energy spectrum of this phase can be found in Ref. \onlinecite{Scherer2012,Scherer2012a}.

\begin{figure}
 \centering
 \includegraphics[width=.95\textwidth]{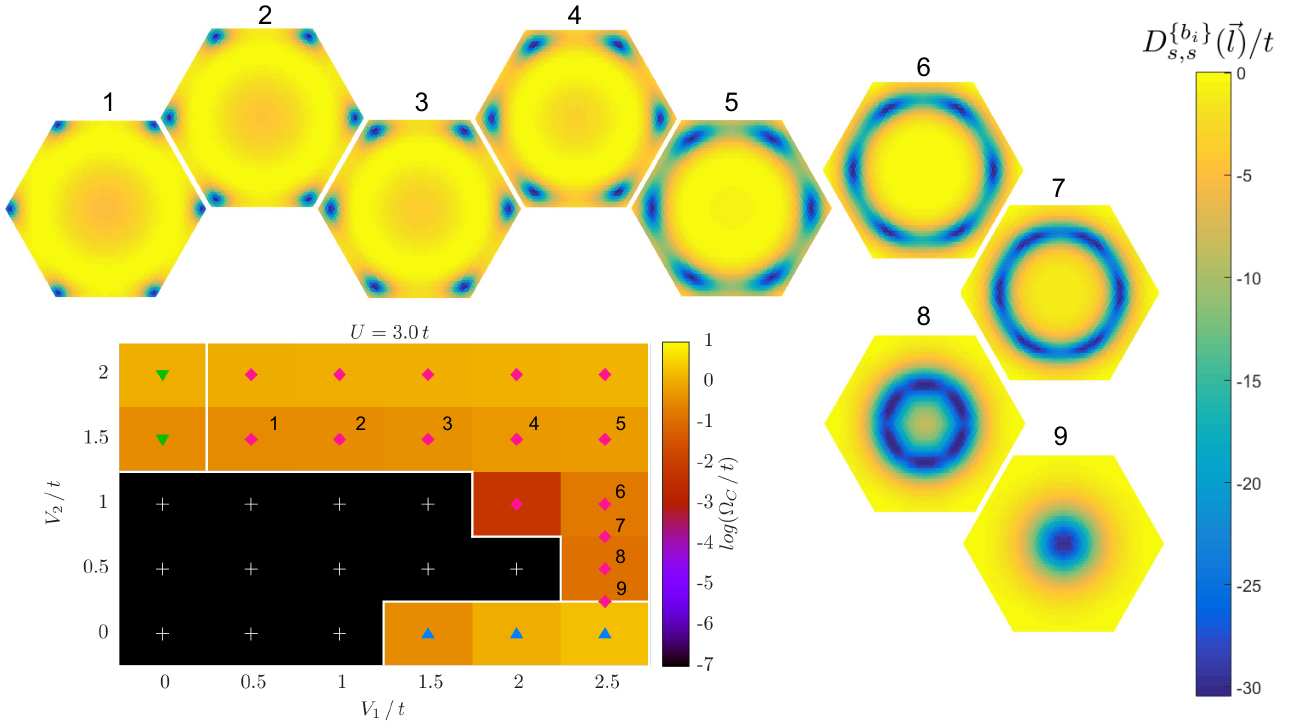}
 \caption{Plots of the $\mathbf l$ dependence of $D^{\{b_i\}}_{s,s}(\mathbf l)$ at the stopping scale for different $V_1/V_2$ ratios. The ordering vectors in plots 1 and 9 are still incommensurate, though very close to $K$ and $\Gamma$ respectively. Since $U$ is subcritical and there is no significant enhancement of magnetic correlations, the behavior of the more physically meaningful $K^{\{b_i\}}_{s,s}(\mathbf l)$ is almost indistinguishable from that of the $D$ propagator above.}
 \label{fig:multiCDW}
\end{figure}

Finally, when both $V_1$ and $V_2$ are supercritical we find incommensurate charge density waves (iCDW's). The system exhibits geometrical frustration since the charge ordering patterns minimizing either first or second nearest neighbor repulsions cannot be realized simultaneously. The ordering vector depends on the ratio $V_1/V_2$, wandering gradually between the two commensurate orderings discussed above as the ratio is modified (see Fig. \ref{fig:multiCDW}). The effective Hamiltonian takes the same form as $H_{\text{CDW}_3}$ but with an ordering vector different from $\mathbf{Q}$. Such incommensurate charge orderings had not yet been observed in previous fRG studies on the honeycomb lattice due to the limited momentum resolution.

\noindent \textit{$\bullet$ Quantum Spin Hall} (QSH) \textit{instability}

A more exotic tendency has caught a lot of interest in recent years. The possibility of a topological Mott insulator\cite{Raghu2008}, an interaction-induced quantum spin hall state, being realized in the honeycomb lattice is currently a source of ongoing debate. In previous results using less refined fRG methods the QSH was triggered by a high enough second neighbor repulsion term. The characteristic correlations for this phase take place in the spin channel for zero wavevector transfer, with the distinctive feature of having an $f$-wave symmetry. It results in the effective Hamiltonian 

\begin{equation}
H_{\text{QSH}}=-\frac{1}{\mathcal{N}} \sum_{o,o'}V_{o,o'}\epsilon_{o}\epsilon_{o'}\mathbf{S}^{o}_{f}\cdot \mathbf{S}^{o'}_{f}
\end{equation}
with $\mathbf{S}^{o}_{f}=\frac{1}{2} \sum_{\mathbf{k},s,s'} f_{\mathbf{k}} \boldsymbol{\sigma}_{s,s'} c^{\dagger}_{\mathbf{k},s,o} c_{\mathbf{k},s',o}$ and $f_{\mathbf{k}}=\sin(\sqrt{3}k_x)-2\sin(\frac{\sqrt{3}k_x}{2})\cos(\frac{3k_y}{2})$. The orbital sign structure is the same as before, but interactions have now an additional $f$-wave modulation that alternates sign between the $K$ and $K'$ points. In a mean-field decoupling of $H_{\text{QSH}}$ an imaginary Kane-Mele order parameter is induced, indicating the formation of an ordered pattern of spin currents with opposite chiralities for the two spin projections.

\subsection{Phase diagrams and critical scales}\label{sec:res1}

\begin{figure}[htb]
  \centering
  \begin{tabular}{@{}c|c@{}}
    \includegraphics[width=.47\textwidth]{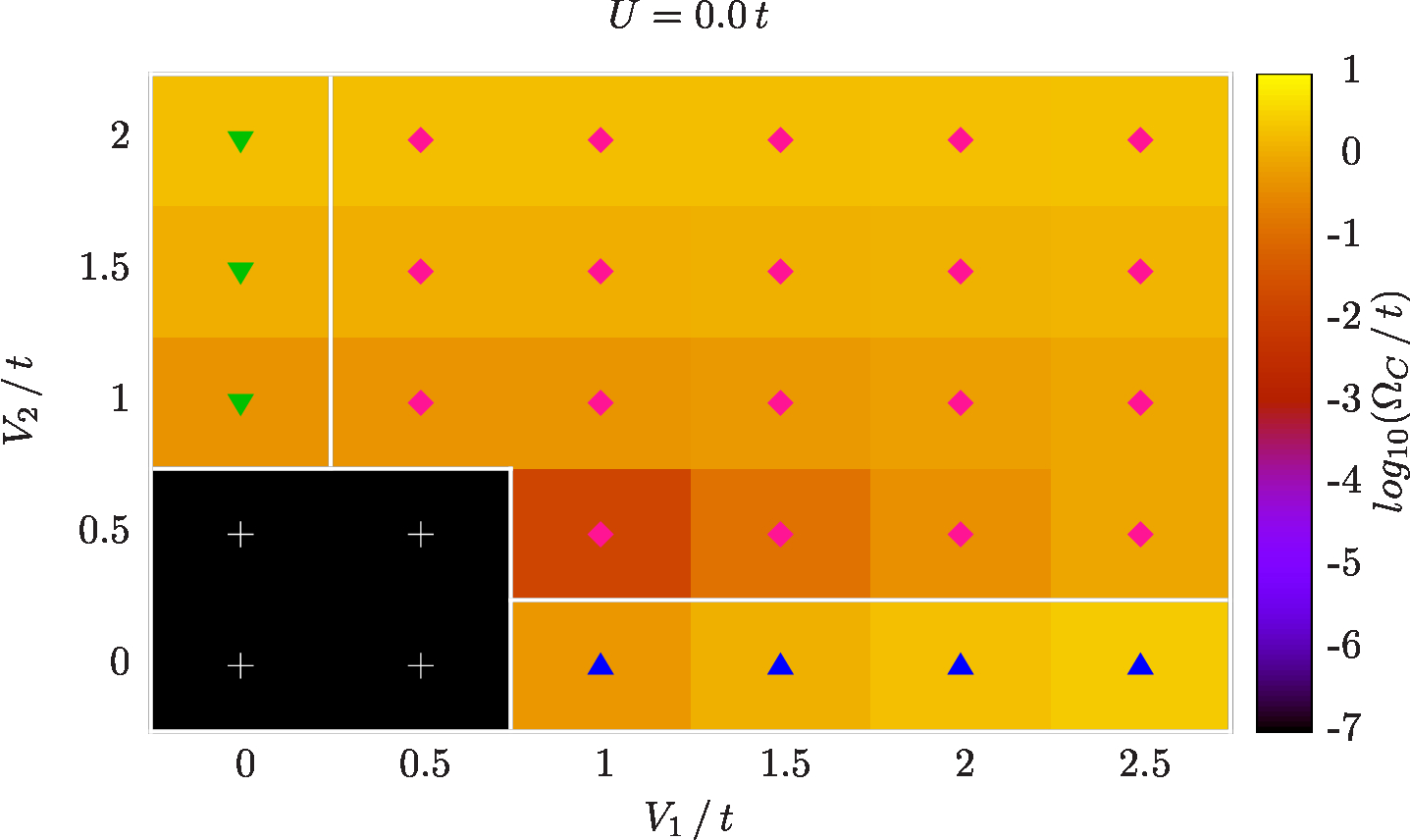}\hspace*{.1cm} &
    \hspace*{.1cm}\includegraphics[width=.47\textwidth]{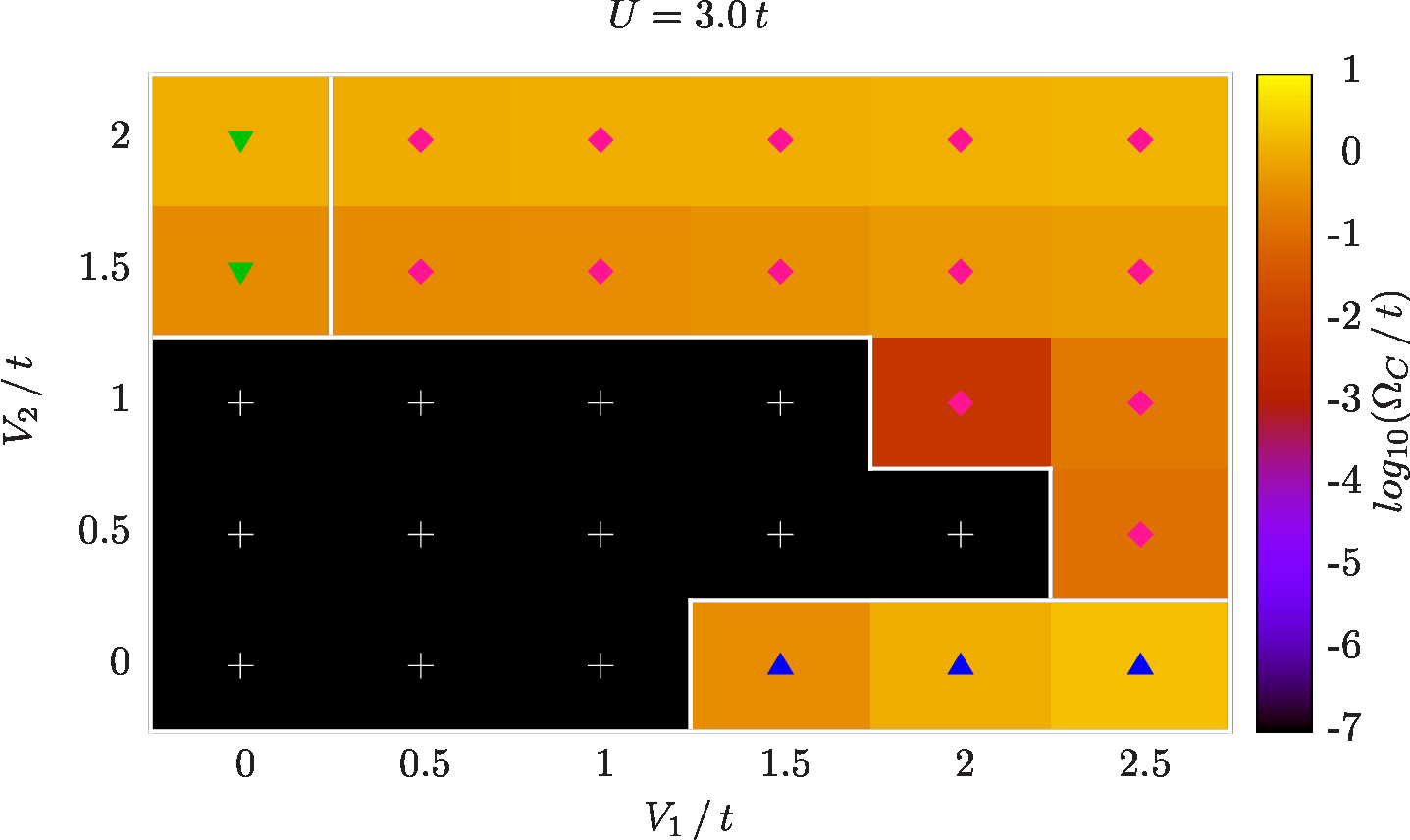}   \\
    \hline
    \\[-3ex]
    \includegraphics[width=.47\textwidth]{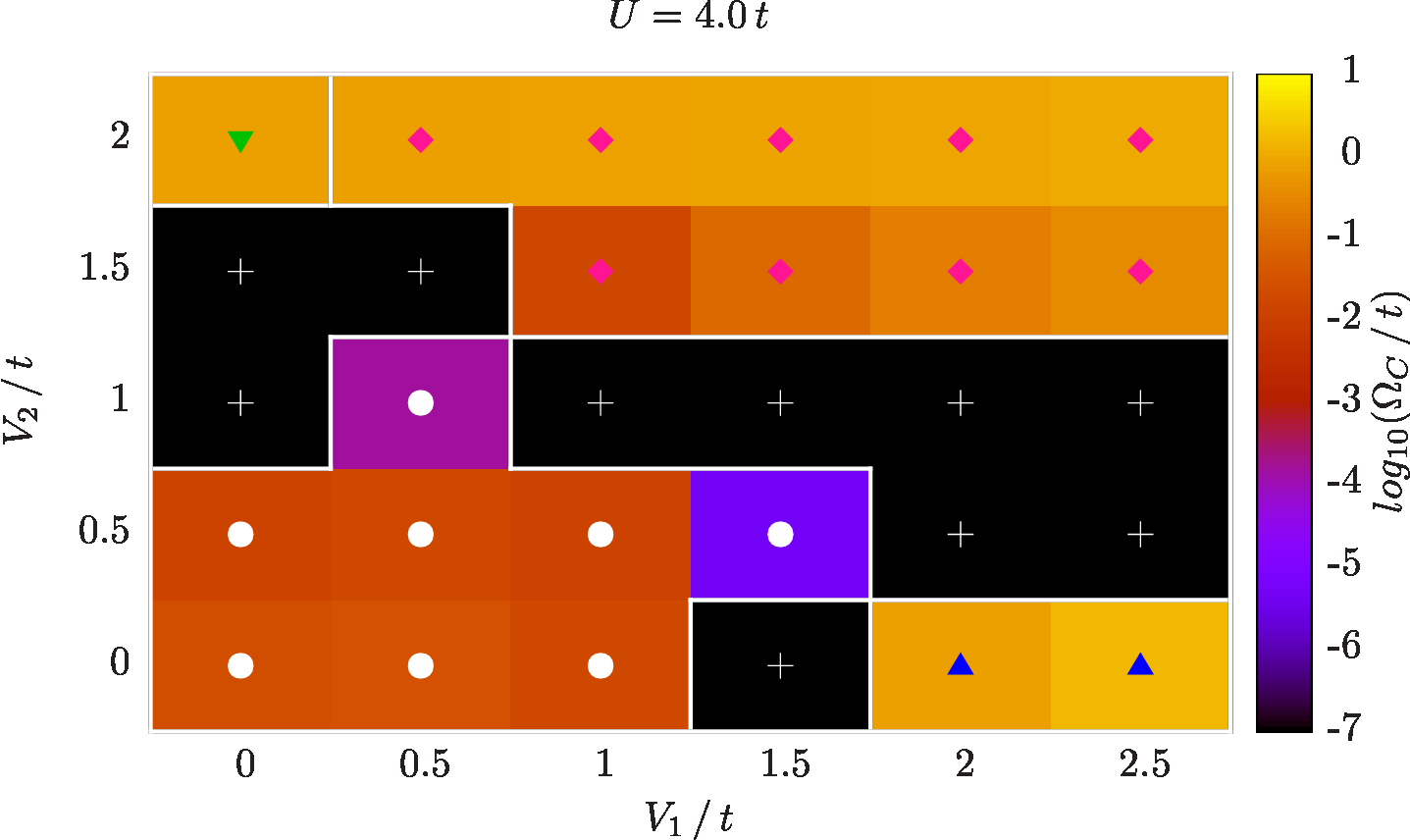}\hspace*{.1cm} &
    \hspace*{.1cm}\includegraphics[width=.47\textwidth]{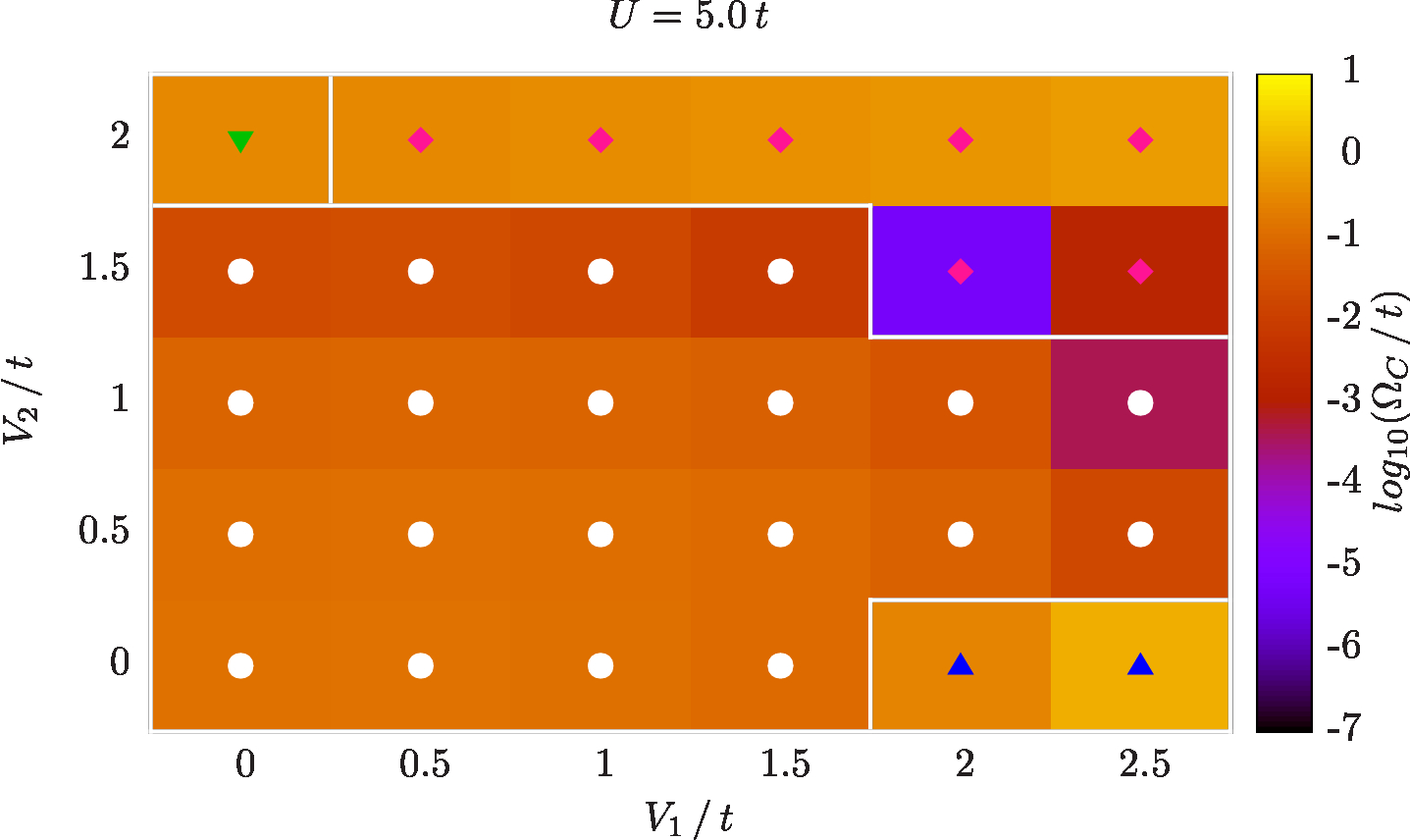}   \\
    \multicolumn{2}{c}{\includegraphics[width=.4\textwidth]{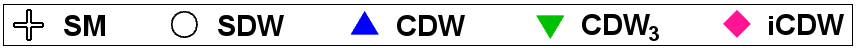}}
  \end{tabular}
  \caption{Dominant instabilities and critical scales for different bare interaction parameters.}
  \label{fig:pd}
\end{figure}

\begin{figure}[htb]
  \centering
  \begin{tabular}{@{}c|c@{}}
    \includegraphics[width=.47\textwidth]{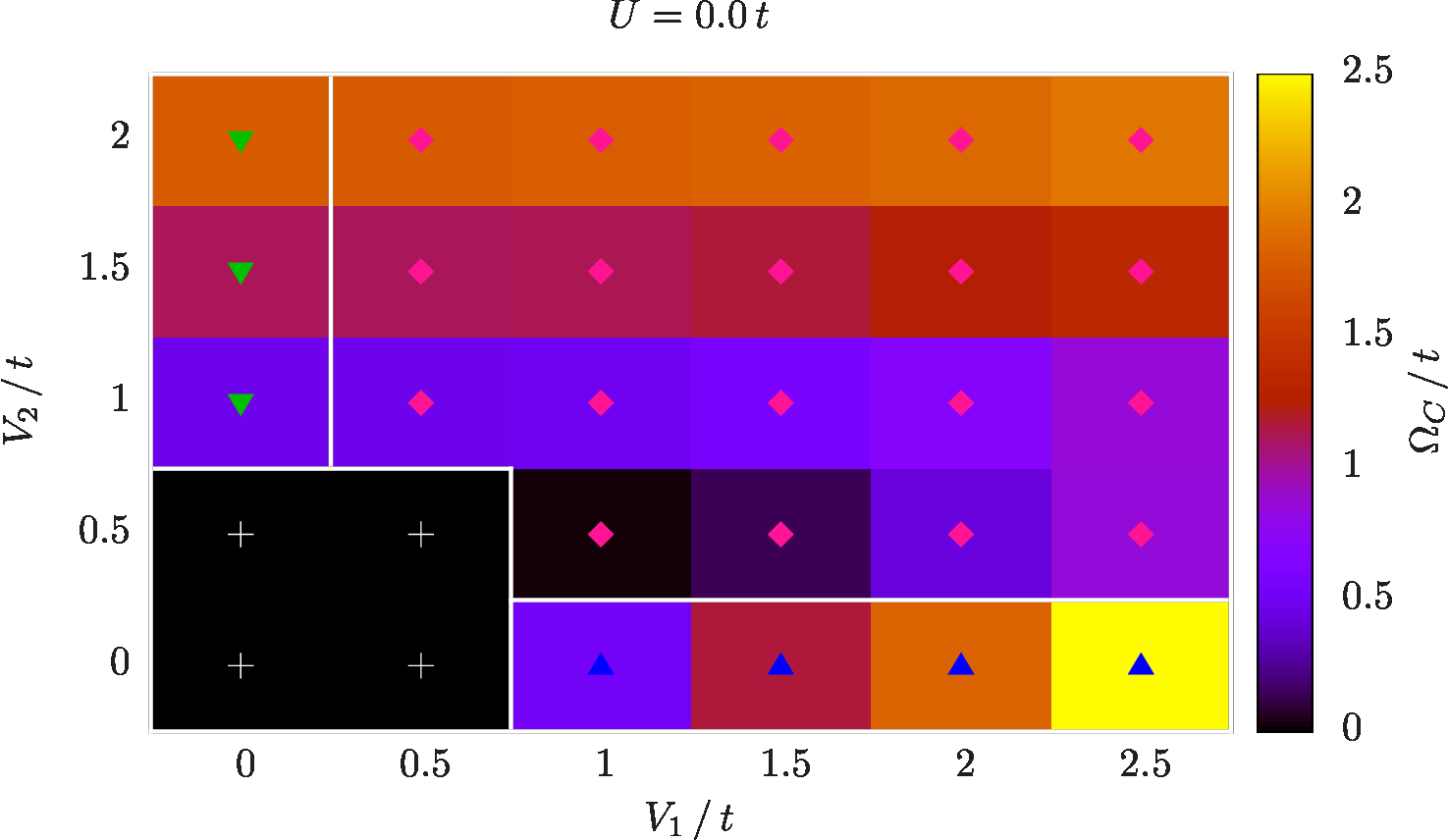}\hspace*{.1cm} &
    \hspace*{.1cm}\includegraphics[width=.47\textwidth]{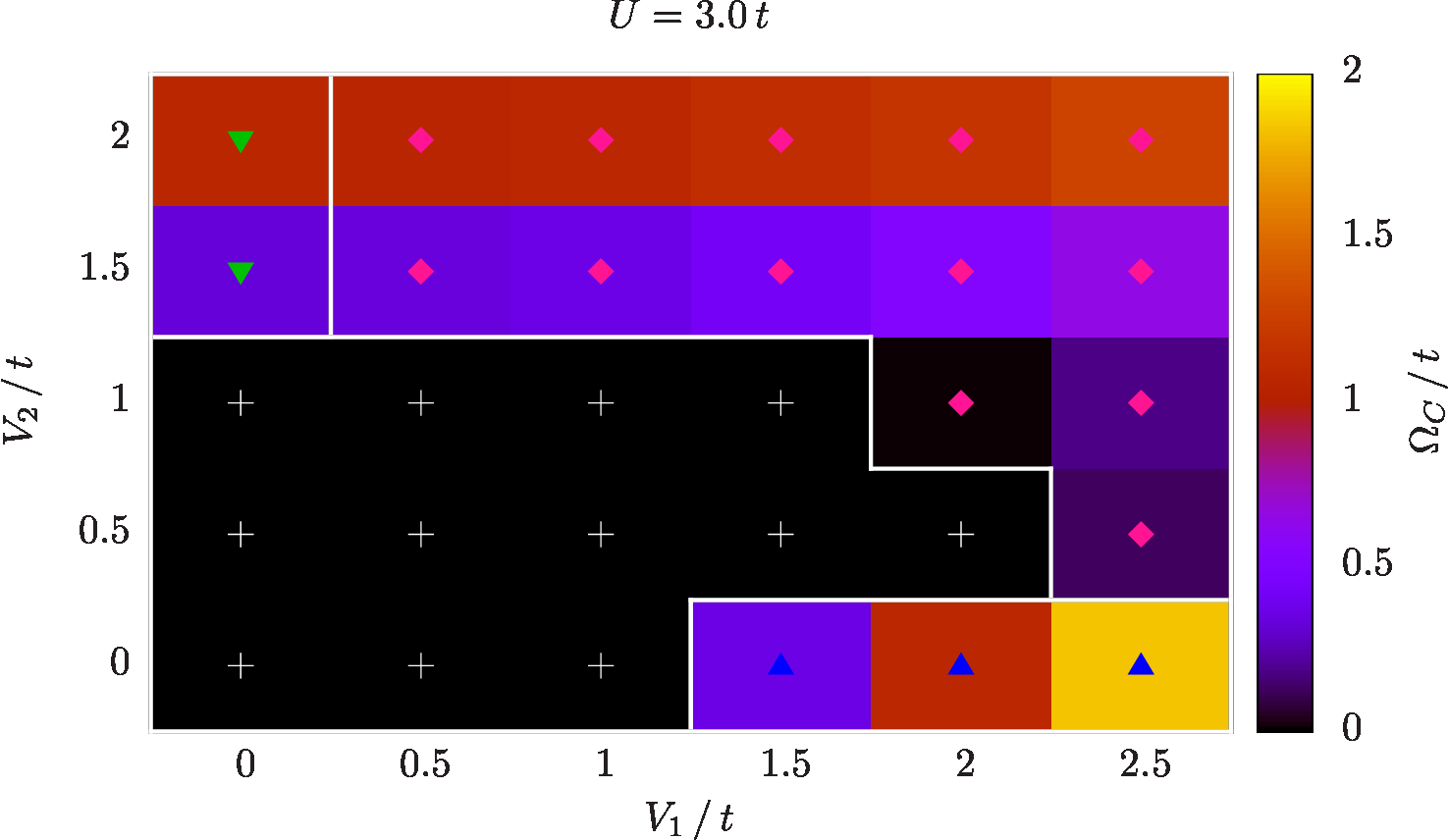}   \\
    \hline
    \\[-3ex]
    \includegraphics[width=.47\textwidth]{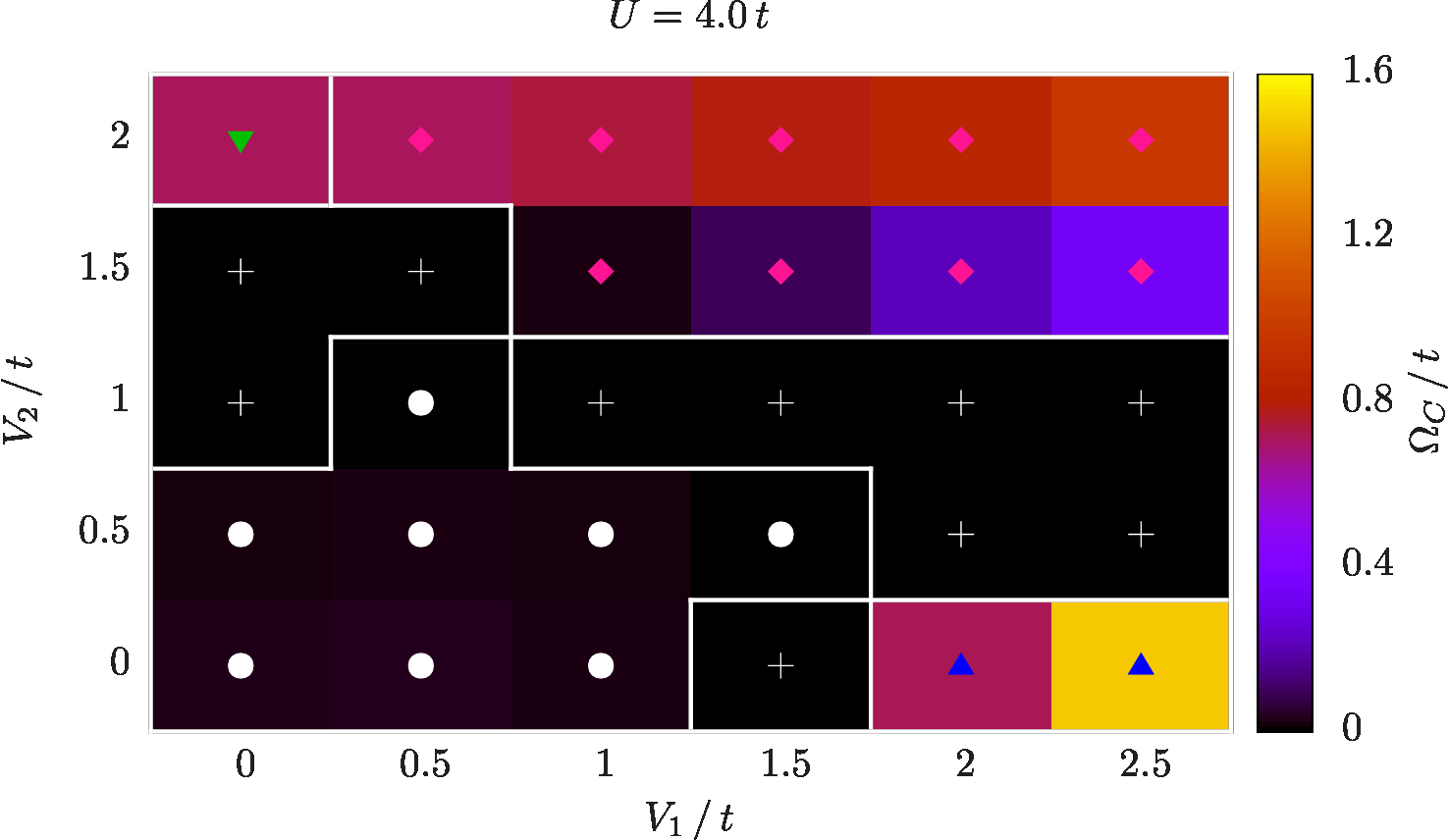}\hspace*{.1cm} &
    \hspace*{.1cm}\includegraphics[width=.47\textwidth]{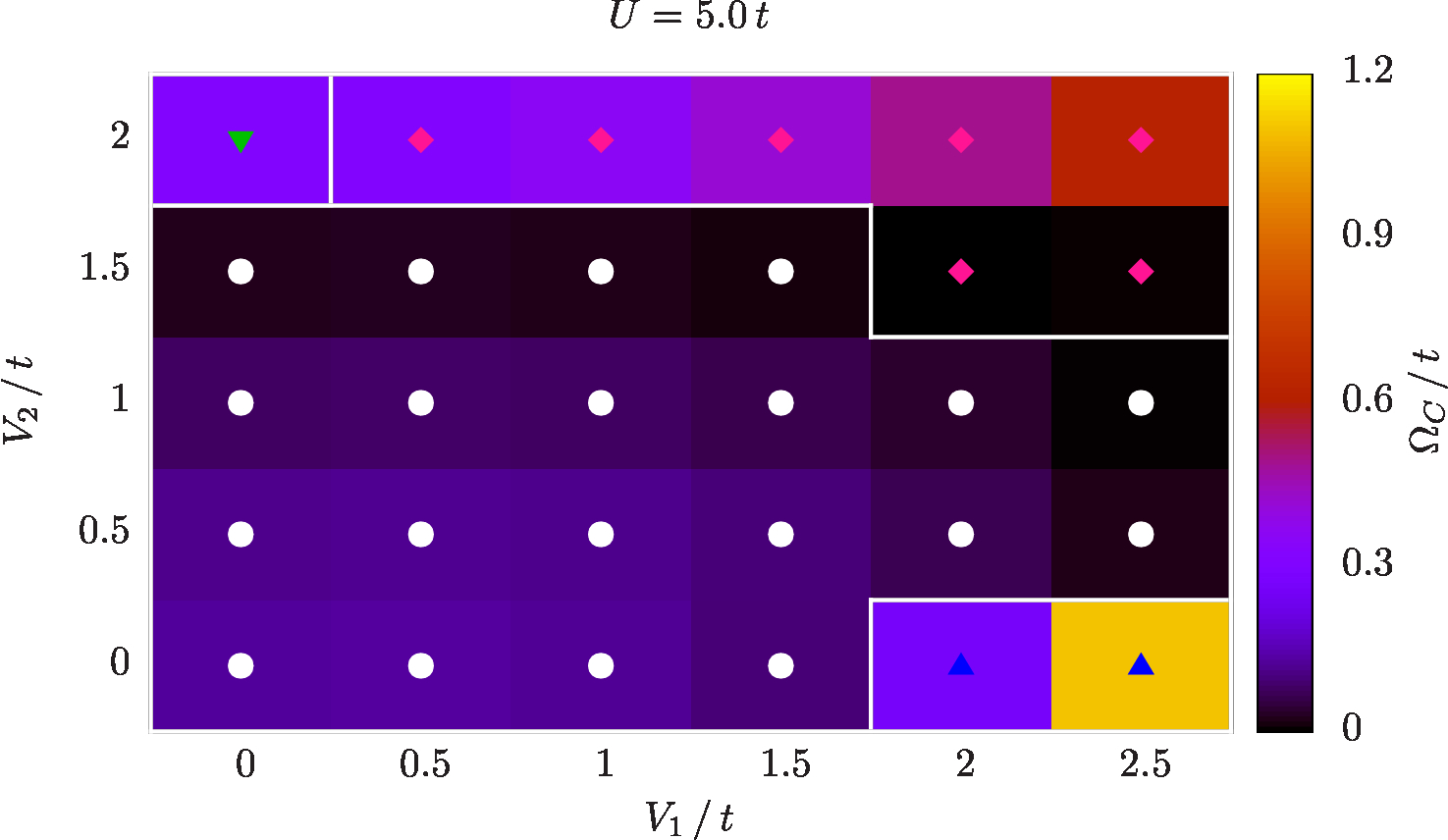}   \\
    \multicolumn{2}{c}{\includegraphics[width=.4\textwidth]{Legend.png}}
  \end{tabular}
  \caption{Phase diagrams of Fig. \ref{fig:pd} with a linear plot of critical scales.}
  \label{fig:pd-lin}
\end{figure}

The phase diagrams obtained are shown in Figs. \ref{fig:pd},\ref{fig:pd-lin}.

In our results, the tendency towards a QSH state is not found to be the dominant instability for any choice of bare interaction parameters. In previous fRG calculations with a pure second neighbor bare coupling, once the value of $V_2$ was chosen to be high enough, the QSH eventually dominated\cite{Raghu2008}. In contrast, we only observe the $\mathrm{CDW}_3$, even up to very high values of $V_2$ where the weak-coupling condition is not fulfilled anymore. In any case, if the ratio $U/V_2$ is small enough, the leading correlations in the spin channel are indeed those responsible for the QSH state. However, their enhancement remains rather modest in comparison with the leading correlations in the charge channel, which are two orders of magnitude bigger at the stopping scale. This scenario of non-dominance for the QSH versus charge order has already been addressed for the QAH in the spinless case with different methods\cite{Jia2013,Daghofer2014,Scherer2015,Capponi2015,Motruk2015}, and more recently for the spinful case\cite{Yanick2016}.

Except for the fact that charge order has taken over the QSH instability, and except for the presence of incommensurate charge order, the phase diagrams are compatible with previous fRG results in terms of the arising tendencies. However, the present method is apparently more sensitive to competition effects, as evidenced by the stronger critical scale variations across the different phases. Though less pronounced, the suppression of critical scales around the boundaries between different tendencies was already captured in previous schemes. Now, even for all three bare coupling parameters taking values which are higher than their individual critical strength, there are regions where the system stays semi-metallic. Despite currently available fRG schemes being certainly not exact, the physical plausibility of a semi-metallic state being stabilized by competition effects makes these results worth considering, although this interesting proposal has yet to be contrasted with other methods.

The resulting instabilities are robust with respect to the inclusion of further shells of form-factors or the use of denser meshes, so long as the locations for ordering vectors in the Brillouin zone are finely discretized. Quantitatively speaking, the truncation of the form-factor basis may influence the critical scales in some regions of the phase diagram. As shown in Ref. \onlinecite{Lichtenstein2016} for the square lattice, the truncation of the form-factor basis has an effect on regions where there is a strong competition between channels. For systems with a vanishing density of states at the Fermi level, the truncation also affects regions around the critical coupling strengths (Fig. \ref{fig:Uc}). Thus the truncation may affect the critical scales at boundaries between magnetic and charge ordered phases, and at boundaries between the semi-metal and ordered phases in general. Including the third shell of form-factors for selected points at such boundaries, some experience an increase in critical scale whereas others find it further suppressed. For instance, for $U=4\,t, \, V_1=1.5\,t, \, V_2=0.5\,t$ the critical scale comes out an order of magnitude higher, but for $U=5\,t, \, V_1=2.5\,t, \, V_2=1.5\,t$ it is an order of magnitude lower than in the $2^{\textrm{nd}}$ shell truncation.

\subsection{Longer ranged interactions and implications for graphene}\label{sec:res2}

Ab-initio interaction parameters for graphene, calculated through the \emph{constrained random phase approximation} (cRPA), are available in the literature\cite{Wehling2011}. For bare coupling parameters up to second nearest neighbor following such cRPA values ($U=3.3\,t, \, V_1=2\,t, \, V_2=1.5\,t$) we find an incommensurate charge density wave instability with a critical scale of $0.47\,t$. Including a third nearest neighbor term with bare coupling strength according to cRPA ($V_3=1.3\,t$) we find no instability. There is no substantial enhancement of any interaction channel at least down to scales of $\Omega = 10^{-9}\,t$. These results are consistent with the experimentally corroborated semi-metallic behavior of undoped single-layer graphene. In previous standard Fermi-surface patching schemes\cite{Scherer2012,Pena2014} the bare $V_3$ interaction term was not included due to limited momentum resolution. The momentum structure of the $V_3$ term is peaked at the $M$ points in the Brillouin zone, and in those works the properly resolved momentum transfers were mainly around the $\Gamma$ and $K,K'$ points but not around $M$ the points. The limited wavevector resolution together with the underestimation of critical coupling strengths had led to a necessary rescaling of ab-initio interaction parameters for graphene. This was justified in order to bring consistency with the lower critical couplings obtained in that scheme, and that way results stayed compatible with the semi-metallic behavior observed in experiments. As shown here, this issue does not arise in a highly wavevector-resolved calculation with enough non-local interaction terms. 

Since Coulomb interactions in pristine graphene at half-filling are not subject to charge screening due to the vanishing density of states at the Fermi level, the influence of further ranged Coulomb repulsion terms is expected to play a role. In particular, such additional coupling terms should induce yet further different charge ordering patterns which may compete with each other. The negative influence of the long-ranged interactions on ordering is known from Quantum Monte Carlo studies\cite{Hohenadler2014}, and the ratio between short and long-range contributions may ultimately decide whether the groundstate is gapped or not\cite{Tang2015}. We plan to analyze that scenario in a forth coming publication. The current scheme is able to resolve bare Coulomb interactions including up to the order of the $1000-th$ nearest neighbor for meshes of a few thousand momentum transfers. Therefore, an attempt to handle a much longer ranged Coulomb tail with fRG is within reach.

\section{CONCLUSION}

In this work we have investigated the effect of improved wavevector resolution and long-range Coulomb interactions on fRG predictions for possible groundstate orderings of electrons in the honeycomb lattice. 

Although the commonly used Fermi-surface patching scheme has brought many insights over the years in capturing the competition of ordering tendencies in an unbiased way, limitations on its predictive power have led to some qualitative discrepancies respect to other theoretical methods and experimental measurements. The TUfRG scheme constitutes a further step to already existing improved parametrizations of the fRG flow\cite{Husemann2009,Eberlein2010,Wang2012,Xiang2013,Eberlein2013,Maier2014,Eberlein2014}, providing an efficient and highly scalable way to refine the Brillouin zone discretization of momentum dependences. There is room for improvement of the scheme, with the most natural extensions being the inclusion of Matsubara frequency dependences and self-energy flows. Nevertheless, the present implementation has already brought some new perspectives on the possible interplay of ordering tendencies in the honeycomb lattice. More specifically, the high wavevector resolution achieved in this work has allowed us to observe a continuous evolution of incommensurate charge orderings for the first time, originating in the frustrated nature of competing interactions in the density channel. When enough competing density-density terms are involved, namely up to the third nearest neighbor, the stability of a semi-metallic state in graphene arises naturally for realistic interaction parameters. Furthermore, our results support the recent findings about the non-dominance of a topological QSH state versus charge order. Using less refined momentum discretizations, the QSH had been found dominant for strong enough second nearest neighbor interactions. Within the present scheme, however, such an instability is strongly suppressed in the whole phase diagram for short-ranged interactions. As in previous results, we do not find any hint for a dominating pairing instability at half-filling.

\section*{acknowledgments}
 We thank Stefan Maier, Timo Reckling, Daniel D. Scherer, Michael M. Scherer, Stefan Wessel and Qiang-Hua Wang for useful discussions. We also thank Daniel Rohe and Edoardo di Napoli for their HPC support. For numerical computation we made use of the DCUHRE routine\cite{Berntsen1991}, the ODEINT libary\cite{Ahnert2011} and the JUBE workflow environment\cite{Luhrs2016}. The authors gratefully acknowledge the computing time granted by JARA-HPC and provided on the supercomputer JURECA at J\"ulich Supercomputing Centre (JSC). This work was financially supported by the DFG research units RTG 1995 and SPP 1459. 

\noindent \section*{APPENDIX: C\lowercase{alculation of projected bare interactions}}

Projectors $\hat P,\hat C, \hat D$ acting over general functions of three momenta $F(\mathbf{k}_1,\mathbf{k}_2, \mathbf{k}_3)$ to respectively bring them into functions of single arguments $\mathbf{l} = \mathbf{k}_1 + \mathbf{k}_2,\, \mathbf{k}_1 - \mathbf{k}_3,\, \mathbf{k}_3 - \mathbf{k}_2$ are defined as

\begin{align}
 \hat P [F]_{m,n}(\mathbf{l}) & =\int \!\!d \mathbf{k} \int \!\! d \mathbf{k}' \,f^*_m(\mathbf{k}) f_n(\mathbf{k}') \, F \!\left(\frac{\mathbf{l}}{2}+\mathbf{k},\frac{\mathbf{l}}{2}-\mathbf{k},\frac{\mathbf{l}}{2}-\mathbf{k}'\right)\,, \notag \\[0.5ex] 
  \hat C [F]_{m,n}(\mathbf{l}) & =\int \!\!d \mathbf{k} \int \!\! d \mathbf{k}' \,f^*_m(\mathbf{k}) f_n(\mathbf{k}') \, F \!\left(\mathbf k+\frac{\mathbf l}{2},\mathbf k'-\frac{\mathbf l}{2},\mathbf k-\frac{\mathbf l}{2}\right)\,, \\[0.5ex] 
 \hat D [F]_{m,n}(\mathbf{l}) & =\int \!\!d \mathbf{k} \int \!\! d \mathbf{k}' \,f^*_m(\mathbf{k}) f_n(\mathbf{k}') \, F \!\left(\mathbf k+\frac{\mathbf l}{2},\mathbf k'-\frac{\mathbf l}{2},\mathbf k'+\frac{\mathbf l}{2}\right)\,. \notag
\end{align}

Fortunately, for the density-density bare interactions of our physical model the four-dimensional integrals above can be split into sums involving two-dimensional integrals. The bare interaction reads

\begin{align}
V^{(0),\,\{b_i\}}(\mathbf k_1,\mathbf k_2, \mathbf k_3) & = \sum_{\{o_i\}} V^{(0),\,\{o_i\}}(\mathbf k_1,\mathbf k_2, \mathbf k_3)\,\hat{T}^{b_1,o_1}_{\mathbf{k}_1}\,\hat{T}^{b_2,o_2}_{\mathbf{k}_2}\left(\hat{T}^{b_3,o_3}_{\mathbf{k}_3}\right)^*  \left(\hat{T}^{b_4,o_4}_{\mathbf{k}_4}\right)^* \,, \notag \\ 
V^{(0),\,\{o_i\}}(\mathbf k_1,\mathbf k_2, \mathbf k_3) & = \sum_{\mathbf{R}_n^{\{o_i\}}} \tilde{V}^{\{o_i\}} \left(\mathbf{R}_n^{\{o_i\}}\right)  e^{i(\mathbf k_3 - \mathbf k_2)\cdot\,\mathbf{R}_n^{\{o_i\}}} \delta_{o_1,o_4}\delta_{o_2,o_3} \,,
\end{align}
where $\hat{T}^{b_i,o_i}_{\mathbf{k}_i}$ are the transformation elements between orbital and band degrees of freedom, $\mathbf{R}_n^{\{o_i\}}$ are intra or inter-orbital $n$-th nearest neighbor bond vectors depending on $\{o_i\}$, and the bare coupling strengths $\tilde{V}^{\{o_i\}} (\mathbf{R}_n^{\{o_i\}})$ are non-zero only for terms up to the furthest neighbor considered, thus the sum over bond vectors contains a finite amount of non-zero terms. Note that for some inter-orbital combinations the interaction involves a conjugate phase $e^{-i(\mathbf k_3 - \mathbf k_2)\cdot\,\mathbf{R}_n^{\{o_i\}}}$, which is accounted for by redefining $\mathbf{R}_n^{\{o_i\}} \rightarrow -\mathbf{R}_n^{\{o_i\}}$, though such bond vectors do not belong to actual lattice positions. The projections take the form

\begin{align}\label{eq:init_proj}
\hat{P}\left[V^{(0),\,\{b_i\}}\right]_{m,n}(\mathbf{l}) & = \sum_{\{o_i\}}\sum_{\mathbf{R}_n^{\{o_i\}}} \tilde{V}^{\{o_i\}} \left(\mathbf{R}_n^{\{o_i\}}\right) \int \!d \mathbf{k} \,f^*_m(\mathbf{k}) \,e^{-i\mathbf{k}\cdot\mathbf{R}_n^{\{o_i\}}} \hat{T}^{b_1,o_1}_{\frac{\mathbf{l}}{2}+\mathbf{k}}\,\hat{T}^{b_2,o_2}_{\frac{\mathbf{l}}{2}-\mathbf{k}} \notag \\ 
& \times \int \!d \mathbf{k}'\, f_n(\mathbf{k}') \, e^{i\mathbf{k}'\cdot\mathbf{R}_n^{\{o_i\}}} \left(\hat{T}^{b_3,o_3}_{\frac{\mathbf{l}}{2}+\mathbf{k}'}\right)^*  \left(\hat{T}^{b_4,o_4}_{\frac{\mathbf{l}}{2}-\mathbf{k}'}\right)^* \,, \notag \\[1.5ex] 
\hat{C}\left[V^{(0),\,\{b_i\}}\right]_{m,n}(\mathbf{l}) & = \sum_{\{o_i\}}\sum_{\mathbf{R}_n^{\{o_i\}}} \tilde{V}^{\{o_i\}} \left(\mathbf{R}_n^{\{o_i\}}\right) \int \!d \mathbf{k} \,f^*_m(\mathbf{k}) \,e^{-i\mathbf{k}\cdot\mathbf{R}_n^{\{o_i\}}} \hat{T}^{b_1,o_1}_{\mathbf{k}+\frac{\mathbf{l}}{2}}\,\left(\hat{T}^{b_4,o_4}_{\mathbf{k}-\frac{\mathbf{l}}{2}}\right)^*  \\ 
& \times \int \!d \mathbf{k}' \,f_n(\mathbf{k}')\, e^{i\mathbf{k}'\cdot\mathbf{R}_n^{\{o_i\}}} \hat{T}^{b_2,o_2}_{\mathbf{k}'-\frac{\mathbf{l}}{2}}  \left(\hat{T}^{b_3,o_3}_{\mathbf{k}'+\frac{\mathbf{l}}{2}}\right)^* \,, \notag \\[1.5ex] 
\hat{D}\left[V^{(0),\,\{b_i\}}\right]_{m,n}(\mathbf{l}) & = \sum_{\{o_i\}}\sum_{\mathbf{R}_n^{\{o_i\}}} \tilde{V}^{\{o_i\}} \left(\mathbf{R}_n^{\{o_i\}}\right) e^{-i\mathbf{l}\cdot\mathbf{R}_n^{\{o_i\}}} \!\int \!d \mathbf{k}\, f^*_m(\mathbf{k}) \, \hat{T}^{b_1,o_1}_{\mathbf{k}+\frac{\mathbf{l}}{2}} \left(\hat{T}^{b_3,o_3}_{\mathbf{k}-\frac{\mathbf{l}}{2}}\right)^*\, \notag \\ 
& \times \int \! d \mathbf{k}'\, f_n(\mathbf{k}') \, \hat{T}^{b_2,o_2}_{\mathbf{k}'-\frac{\mathbf{l}}{2}}   \left(\hat{T}^{b_4,o_4}_{\mathbf{k}'+\frac{\mathbf{l}}{2}}\right)^* \,. \notag
\end{align}
The unitary transformation $\hat{\mathbf{T}}_{\mathbf{k}}$ is chosen as

\begin{align}\label{eq:T_k1}
 \hat{\mathbf{T}}_{\mathbf{k}} & = \frac{1}{\sqrt{2}}
\begin{pmatrix}
\frac{h(\mathbf{k})}{\left|h(\mathbf{k})\right|} & -1 \\
1 & \frac{h^*(\mathbf{k})}{\left|h(\mathbf{k})\right|}
\end{pmatrix} \\
h(\mathbf{k}) & = \sum_{\boldsymbol{\delta}} e^{i \mathbf{k} \cdot \boldsymbol{\delta}} \notag
\end{align}
where $\boldsymbol{\delta}=\{\mathbf{R}_1^{\mathrm{ABAB}}\}$ are the nearest neighbor bond vectors. Since transformation elements satisfy $\left(\hat{T}^{b,o}_{\mathbf{k}}\right)^*=\hat{T}^{b,o}_{-\mathbf{k}}$, one just needs to calculate the integral

\begin{equation}\label{eq:init_proj_i}
 \int \!d \mathbf{k} \,f_m(\mathbf{k}) \,e^{i\mathbf{k}\cdot\mathbf{R}_n^{\{o_i\}}} \hat{T}^{b,o}_{\mathbf{k}-\frac{\mathbf{l}}{2}}\,\left(\hat{T}^{b',o'}_{\mathbf{k}+\frac{\mathbf{l}}{2}}\right)^*
\end{equation}
for all $f_m(\mathbf{k})$, $\mathbf{l}$, $b$, $b'$, $o$, $o'$ and corresponding $\mathbf{R}_n^{\{o_i\}}$ to construct the result of all three projections, where the $D$ channel integrands amount to the special case involving just the on-site bond vector.

As a technical side note, it is worth discussing the implications of the choice of Bloch basis\cite{Maier2013a} in our calculation. By choice of Bloch basis, we are referring to the U(1) invariance of the electronic structure under $\mathbf{k}$-local phase transformations of the fields and vertex functions, which also influences the form taken by $\hat{\mathbf{T}}_{\mathbf{k}}$. One would be tempted to work in the so-called \textit{proper gauge} or \textit{proper basis}\cite{Maier2013a}, where the local phase is chosen so that all objects inside the integrals in Eq. (\ref{eq:init_proj}) have the periodicity of the reciprocal lattice, and thus the integrands are smooth. However, in this basis, the otherwise trivial behavior of $\hat{\mathbf{T}}_{\mathbf{k}}$ and inter-orbital coupling components under point group symmetries becomes non-trivial. In turn, the physical interpretation of the instability analysis in terms of irreducible representations of the lattice point group is obscured. We stick to the standard basis instead, even though one must deal with discontinuities in the integrands of Eq. (\ref{eq:init_proj}) due to the back-folding of non-periodic functions into the first Brillouin zone. 

Furthermore, one is also free to do U(1) transformations on the individual eigenvectors composing $\hat{\mathbf{T}}_{\mathbf{k}}$, redistributing the weight among sublattices. For example, an equally valid choice is

\begin{align}\label{eq:T_k2}
 \hat{\mathbf{T}}_{\mathbf{k}} & = \frac{1}{\sqrt{2}}
\begin{pmatrix}
\frac{h(\mathbf{k})}{\left|h(\mathbf{k})\right|} & -1 \\
\frac{h(\mathbf{k})}{\left|h(\mathbf{k})\right|} & 1
\end{pmatrix} \\
h(\mathbf{k}) & = \sum_{\boldsymbol{\delta}} e^{i \mathbf{k} \cdot \boldsymbol{\delta}} \notag
\end{align}

This is the choice referred to in section \ref{sec:res0} when discussing critical onsite interaction strengths. It is deemed to be a more appropiate choice than \ref{eq:T_k1}, since the resulting integrands in \ref{eq:init_proj} are better behaved. Tracking the difference between these choices in the analytical expressions \ref{eq:init_proj} is complicated, but positive numerical implications are manifest in shorter computation times for the projection of bare interactions, and much faster convergence of $U_C$ values respect to the number of form-factors or BZ mesh points. Since it also delivers a $U_C$ value which is more compatible with other fRG works, we suggest \ref{eq:T_k2} as a reliable choice for the orbital to band transformation within our scheme.


\nocite{*}
\bibliography{bibliography}

\begin{thebibliography}{52}%
\makeatletter
\providecommand \@ifxundefined [1]{%
 \@ifx{#1\undefined}
}%
\providecommand \@ifnum [1]{%
 \ifnum #1\expandafter \@firstoftwo
 \else \expandafter \@secondoftwo
 \fi
}%
\providecommand \@ifx [1]{%
 \ifx #1\expandafter \@firstoftwo
 \else \expandafter \@secondoftwo
 \fi
}%
\providecommand \natexlab [1]{#1}%
\providecommand \enquote  [1]{``#1''}%
\providecommand \bibnamefont  [1]{#1}%
\providecommand \bibfnamefont [1]{#1}%
\providecommand \citenamefont [1]{#1}%
\providecommand \href@noop [0]{\@secondoftwo}%
\providecommand \href [0]{\begingroup \@sanitize@url \@href}%
\providecommand \@href[1]{\@@startlink{#1}\@@href}%
\providecommand \@@href[1]{\endgroup#1\@@endlink}%
\providecommand \@sanitize@url [0]{\catcode `\\12\catcode `\$12\catcode
  `\&12\catcode `\#12\catcode `\^12\catcode `\_12\catcode `\%12\relax}%
\providecommand \@@startlink[1]{}%
\providecommand \@@endlink[0]{}%
\providecommand \url  [0]{\begingroup\@sanitize@url \@url }%
\providecommand \@url [1]{\endgroup\@href {#1}{\urlprefix }}%
\providecommand \urlprefix  [0]{URL }%
\providecommand \Eprint [0]{\href }%
\providecommand \doibase [0]{http://dx.doi.org/}%
\providecommand \selectlanguage [0]{\@gobble}%
\providecommand \bibinfo  [0]{\@secondoftwo}%
\providecommand \bibfield  [0]{\@secondoftwo}%
\providecommand \translation [1]{[#1]}%
\providecommand \BibitemOpen [0]{}%
\providecommand \bibitemStop [0]{}%
\providecommand \bibitemNoStop [0]{.\EOS\space}%
\providecommand \EOS [0]{\spacefactor3000\relax}%
\providecommand \BibitemShut  [1]{\csname bibitem#1\endcsname}%
\let\auto@bib@innerbib\@empty
\bibitem [{\citenamefont {{Geim}}\ and\ \citenamefont
  {{Novoselov}}(2007)}]{Geim2007}%
  \BibitemOpen
  \bibfield  {author} {\bibinfo {author} {\bibfnamefont {A.~K.}\ \bibnamefont
  {{Geim}}}\ and\ \bibinfo {author} {\bibfnamefont {K.~S.}\ \bibnamefont
  {{Novoselov}}},\ }\href@noop {} {\bibfield  {journal} {\bibinfo  {journal}
  {Nat. Mater}\ }\textbf {\bibinfo {volume} {183}} (\bibinfo {year}
  {2007})}\BibitemShut {NoStop}%
\bibitem [{\citenamefont {{Castro Neto}}\ \emph {et~al.}(2009)\citenamefont
  {{Castro Neto}}, \citenamefont {{Guinea}}, \citenamefont {{Peres}},
  \citenamefont {{Novoselov}},\ and\ \citenamefont {{Geim}}}]{CastroNeto2009}%
  \BibitemOpen
  \bibfield  {author} {\bibinfo {author} {\bibfnamefont {A.~H.}\ \bibnamefont
  {{Castro Neto}}}, \bibinfo {author} {\bibfnamefont {F.}~\bibnamefont
  {{Guinea}}}, \bibinfo {author} {\bibfnamefont {N.~M.~R.}\ \bibnamefont
  {{Peres}}}, \bibinfo {author} {\bibfnamefont {K.~S.}\ \bibnamefont
  {{Novoselov}}}, \ and\ \bibinfo {author} {\bibfnamefont {A.~K.}\ \bibnamefont
  {{Geim}}},\ }\href {\doibase 10.1103/RevModPhys.81.109} {\bibfield  {journal}
  {\bibinfo  {journal} {Reviews of Modern Physics}\ }\textbf {\bibinfo {volume}
  {81}},\ \bibinfo {pages} {109} (\bibinfo {year} {2009})}\BibitemShut
  {NoStop}%
\bibitem [{\citenamefont {{Sorella}}\ and\ \citenamefont
  {{Tosatti}}(1992)}]{Sorella1992}%
  \BibitemOpen
  \bibfield  {author} {\bibinfo {author} {\bibfnamefont {S.}~\bibnamefont
  {{Sorella}}}\ and\ \bibinfo {author} {\bibfnamefont {E.}~\bibnamefont
  {{Tosatti}}},\ }\href {\doibase 10.1209/0295-5075/19/8/007} {\bibfield
  {journal} {\bibinfo  {journal} {EPL (Europhysics Letters)}\ }\textbf
  {\bibinfo {volume} {19}},\ \bibinfo {pages} {699} (\bibinfo {year}
  {1992})}\BibitemShut {NoStop}%
\bibitem [{\citenamefont {{Herbut}}(2006)}]{Herbut2006}%
  \BibitemOpen
  \bibfield  {author} {\bibinfo {author} {\bibfnamefont {I.~F.}\ \bibnamefont
  {{Herbut}}},\ }\href {\doibase 10.1103/PhysRevLett.97.146401} {\bibfield
  {journal} {\bibinfo  {journal} {Physical Review Letters}\ }\textbf {\bibinfo
  {volume} {97}},\ \bibinfo {eid} {146401} (\bibinfo {year}
  {2006})}\BibitemShut {NoStop}%
\bibitem [{\citenamefont {Vafek}(2007)}]{Vafek2007}%
  \BibitemOpen
  \bibfield  {author} {\bibinfo {author} {\bibfnamefont {O.}~\bibnamefont
  {Vafek}},\ }\href {\doibase 10.1103/PhysRevLett.98.216401} {\bibfield
  {journal} {\bibinfo  {journal} {Phys. Rev. Lett.}\ }\textbf {\bibinfo
  {volume} {98}},\ \bibinfo {pages} {216401} (\bibinfo {year}
  {2007})}\BibitemShut {NoStop}%
\bibitem [{\citenamefont {{Uchoa}}\ and\ \citenamefont {{Castro
  Neto}}(2007)}]{Uchoa2007}%
  \BibitemOpen
  \bibfield  {author} {\bibinfo {author} {\bibfnamefont {B.}~\bibnamefont
  {{Uchoa}}}\ and\ \bibinfo {author} {\bibfnamefont {A.~H.}\ \bibnamefont
  {{Castro Neto}}},\ }\href {\doibase 10.1103/PhysRevLett.98.146801} {\bibfield
   {journal} {\bibinfo  {journal} {Physical Review Letters}\ }\textbf {\bibinfo
  {volume} {98}},\ \bibinfo {eid} {146801} (\bibinfo {year}
  {2007})}\BibitemShut {NoStop}%
\bibitem [{\citenamefont {Honerkamp}(2008)}]{Honerkamp2008}%
  \BibitemOpen
  \bibfield  {author} {\bibinfo {author} {\bibfnamefont {C.}~\bibnamefont
  {Honerkamp}},\ }\href {\doibase 10.1103/PhysRevLett.100.146404} {\bibfield
  {journal} {\bibinfo  {journal} {Phys. Rev. Lett.}\ }\textbf {\bibinfo
  {volume} {100}},\ \bibinfo {pages} {146404} (\bibinfo {year}
  {2008})}\BibitemShut {NoStop}%
\bibitem [{\citenamefont {Herbut}\ \emph
  {et~al.}(2009{\natexlab{a}})\citenamefont {Herbut}, \citenamefont
  {Juri\ifmmode \check{c}\else \v{c}\fi{}i\ifmmode~\acute{c}\else \'{c}\fi{}},\
  and\ \citenamefont {Vafek}}]{Herbut2009}%
  \BibitemOpen
  \bibfield  {author} {\bibinfo {author} {\bibfnamefont {I.~F.}\ \bibnamefont
  {Herbut}}, \bibinfo {author} {\bibfnamefont {V.}~\bibnamefont {Juri\ifmmode
  \check{c}\else \v{c}\fi{}i\ifmmode~\acute{c}\else \'{c}\fi{}}}, \ and\
  \bibinfo {author} {\bibfnamefont {O.}~\bibnamefont {Vafek}},\ }\href
  {\doibase 10.1103/PhysRevB.80.075432} {\bibfield  {journal} {\bibinfo
  {journal} {Phys. Rev. B}\ }\textbf {\bibinfo {volume} {80}},\ \bibinfo
  {pages} {075432} (\bibinfo {year} {2009}{\natexlab{a}})}\BibitemShut
  {NoStop}%
\bibitem [{\citenamefont {Herbut}\ \emph
  {et~al.}(2009{\natexlab{b}})\citenamefont {Herbut}, \citenamefont
  {Juri\ifmmode \check{c}\else \v{c}\fi{}i\ifmmode~\acute{c}\else \'{c}\fi{}},\
  and\ \citenamefont {Roy}}]{Herbut2009a}%
  \BibitemOpen
  \bibfield  {author} {\bibinfo {author} {\bibfnamefont {I.~F.}\ \bibnamefont
  {Herbut}}, \bibinfo {author} {\bibfnamefont {V.}~\bibnamefont {Juri\ifmmode
  \check{c}\else \v{c}\fi{}i\ifmmode~\acute{c}\else \'{c}\fi{}}}, \ and\
  \bibinfo {author} {\bibfnamefont {B.}~\bibnamefont {Roy}},\ }\href {\doibase
  10.1103/PhysRevB.79.085116} {\bibfield  {journal} {\bibinfo  {journal} {Phys.
  Rev. B}\ }\textbf {\bibinfo {volume} {79}},\ \bibinfo {pages} {085116}
  (\bibinfo {year} {2009}{\natexlab{b}})}\BibitemShut {NoStop}%
\bibitem [{\citenamefont {{Meng}}\ \emph {et~al.}(2010)\citenamefont {{Meng}},
  \citenamefont {{Lang}}, \citenamefont {{Wessel}}, \citenamefont {{Assaad}},\
  and\ \citenamefont {{Muramatsu}}}]{Meng2010}%
  \BibitemOpen
  \bibfield  {author} {\bibinfo {author} {\bibfnamefont {Z.~Y.}\ \bibnamefont
  {{Meng}}}, \bibinfo {author} {\bibfnamefont {T.~C.}\ \bibnamefont {{Lang}}},
  \bibinfo {author} {\bibfnamefont {S.}~\bibnamefont {{Wessel}}}, \bibinfo
  {author} {\bibfnamefont {F.~F.}\ \bibnamefont {{Assaad}}}, \ and\ \bibinfo
  {author} {\bibfnamefont {A.}~\bibnamefont {{Muramatsu}}},\ }\href {\doibase
  10.1038/nature08942} {\bibfield  {journal} {\bibinfo  {journal} {\nat}\
  }\textbf {\bibinfo {volume} {464}},\ \bibinfo {pages} {847} (\bibinfo {year}
  {2010})}\BibitemShut {NoStop}%
\bibitem [{\citenamefont {{Sorella}}\ \emph {et~al.}(2012)\citenamefont
  {{Sorella}}, \citenamefont {{Otsuka}},\ and\ \citenamefont
  {{Yunoki}}}]{Sorella2012}%
  \BibitemOpen
  \bibfield  {author} {\bibinfo {author} {\bibfnamefont {S.}~\bibnamefont
  {{Sorella}}}, \bibinfo {author} {\bibfnamefont {Y.}~\bibnamefont {{Otsuka}}},
  \ and\ \bibinfo {author} {\bibfnamefont {S.}~\bibnamefont {{Yunoki}}},\
  }\href {\doibase 10.1038/srep00992} {\bibfield  {journal} {\bibinfo
  {journal} {Scientific Reports}\ }\textbf {\bibinfo {volume} {2}},\ \bibinfo
  {eid} {992} (\bibinfo {year} {2012})}\BibitemShut {NoStop}%
\bibitem [{\citenamefont {{Ulybyshev}}\ \emph {et~al.}(2013)\citenamefont
  {{Ulybyshev}}, \citenamefont {{Buividovich}}, \citenamefont {{Katsnelson}},\
  and\ \citenamefont {{Polikarpov}}}]{Ulybishev2013}%
  \BibitemOpen
  \bibfield  {author} {\bibinfo {author} {\bibfnamefont {M.~V.}\ \bibnamefont
  {{Ulybyshev}}}, \bibinfo {author} {\bibfnamefont {P.~V.}\ \bibnamefont
  {{Buividovich}}}, \bibinfo {author} {\bibfnamefont {M.~I.}\ \bibnamefont
  {{Katsnelson}}}, \ and\ \bibinfo {author} {\bibfnamefont {M.~I.}\
  \bibnamefont {{Polikarpov}}},\ }\href {\doibase
  10.1103/PhysRevLett.111.056801} {\bibfield  {journal} {\bibinfo  {journal}
  {Physical Review Letters}\ }\textbf {\bibinfo {volume} {111}},\ \bibinfo
  {eid} {056801} (\bibinfo {year} {2013})}\BibitemShut {NoStop}%
\bibitem [{\citenamefont {{Classen}}\ \emph {et~al.}(2014)\citenamefont
  {{Classen}}, \citenamefont {{Scherer}},\ and\ \citenamefont
  {{Honerkamp}}}]{Classen2014}%
  \BibitemOpen
  \bibfield  {author} {\bibinfo {author} {\bibfnamefont {L.}~\bibnamefont
  {{Classen}}}, \bibinfo {author} {\bibfnamefont {M.~M.}\ \bibnamefont
  {{Scherer}}}, \ and\ \bibinfo {author} {\bibfnamefont {C.}~\bibnamefont
  {{Honerkamp}}},\ }\href {\doibase 10.1103/PhysRevB.90.035122} {\bibfield
  {journal} {\bibinfo  {journal} {\prb}\ }\textbf {\bibinfo {volume} {90}},\
  \bibinfo {eid} {035122} (\bibinfo {year} {2014})}\BibitemShut {NoStop}%
\bibitem [{\citenamefont {{Hohenadler}}\ \emph {et~al.}(2014)\citenamefont
  {{Hohenadler}}, \citenamefont {{Parisen Toldin}}, \citenamefont {{Herbut}},\
  and\ \citenamefont {{Assaad}}}]{Hohenadler2014}%
  \BibitemOpen
  \bibfield  {author} {\bibinfo {author} {\bibfnamefont {M.}~\bibnamefont
  {{Hohenadler}}}, \bibinfo {author} {\bibfnamefont {F.}~\bibnamefont {{Parisen
  Toldin}}}, \bibinfo {author} {\bibfnamefont {I.~F.}\ \bibnamefont
  {{Herbut}}}, \ and\ \bibinfo {author} {\bibfnamefont {F.~F.}\ \bibnamefont
  {{Assaad}}},\ }\href {\doibase 10.1103/PhysRevB.90.085146} {\bibfield
  {journal} {\bibinfo  {journal} {\prb}\ }\textbf {\bibinfo {volume} {90}},\
  \bibinfo {eid} {085146} (\bibinfo {year} {2014})}\BibitemShut {NoStop}%
\bibitem [{\citenamefont {{Golor}}\ and\ \citenamefont
  {{Wessel}}(2015)}]{Golor2015}%
  \BibitemOpen
  \bibfield  {author} {\bibinfo {author} {\bibfnamefont {M.}~\bibnamefont
  {{Golor}}}\ and\ \bibinfo {author} {\bibfnamefont {S.}~\bibnamefont
  {{Wessel}}},\ }\href {\doibase 10.1103/PhysRevB.92.195154} {\bibfield
  {journal} {\bibinfo  {journal} {\prb}\ }\textbf {\bibinfo {volume} {92}},\
  \bibinfo {eid} {195154} (\bibinfo {year} {2015})}\BibitemShut {NoStop}%
\bibitem [{\citenamefont {{Tang}}\ \emph {et~al.}(2015)\citenamefont {{Tang}},
  \citenamefont {{Laksono}}, \citenamefont {{Rodrigues}}, \citenamefont
  {{Sengupta}}, \citenamefont {{Assaad}},\ and\ \citenamefont
  {{Adam}}}]{Tang2015}%
  \BibitemOpen
  \bibfield  {author} {\bibinfo {author} {\bibfnamefont {H.-K.}\ \bibnamefont
  {{Tang}}}, \bibinfo {author} {\bibfnamefont {E.}~\bibnamefont {{Laksono}}},
  \bibinfo {author} {\bibfnamefont {J.~N.~B.}\ \bibnamefont {{Rodrigues}}},
  \bibinfo {author} {\bibfnamefont {P.}~\bibnamefont {{Sengupta}}}, \bibinfo
  {author} {\bibfnamefont {F.~F.}\ \bibnamefont {{Assaad}}}, \ and\ \bibinfo
  {author} {\bibfnamefont {S.}~\bibnamefont {{Adam}}},\ }\href {\doibase
  10.1103/PhysRevLett.115.186602} {\bibfield  {journal} {\bibinfo  {journal}
  {Physical Review Letters}\ }\textbf {\bibinfo {volume} {115}},\ \bibinfo
  {eid} {186602} (\bibinfo {year} {2015})}\BibitemShut {NoStop}%
\bibitem [{\citenamefont {{Lichtenstein}}\ \emph
  {et~al.}(2016{\natexlab{a}})\citenamefont {{Lichtenstein}}, \citenamefont
  {{S{\'a}nchez de la Pe{\~n}a}}, \citenamefont {{Rohe}}, \citenamefont {{Di
  Napoli}}, \citenamefont {{Honerkamp}},\ and\ \citenamefont
  {{Maier}}}]{Lichtenstein2016}%
  \BibitemOpen
  \bibfield  {author} {\bibinfo {author} {\bibfnamefont {J.}~\bibnamefont
  {{Lichtenstein}}}, \bibinfo {author} {\bibfnamefont {D.}~\bibnamefont
  {{S{\'a}nchez de la Pe{\~n}a}}}, \bibinfo {author} {\bibfnamefont
  {D.}~\bibnamefont {{Rohe}}}, \bibinfo {author} {\bibfnamefont
  {E.}~\bibnamefont {{Di Napoli}}}, \bibinfo {author} {\bibfnamefont
  {C.}~\bibnamefont {{Honerkamp}}}, \ and\ \bibinfo {author} {\bibfnamefont
  {S.~A.}\ \bibnamefont {{Maier}}},\ }\href@noop {} {\bibfield  {journal}
  {\bibinfo  {journal} {ArXiv e-prints}\ } (\bibinfo {year}
  {2016}{\natexlab{a}})},\ \Eprint {http://arxiv.org/abs/1604.06296}
  {arXiv:1604.06296 [cond-mat.str-el]} \BibitemShut {NoStop}%
\bibitem [{\citenamefont {Husemann}\ and\ \citenamefont
  {Salmhofer}(2009)}]{Husemann2009}%
  \BibitemOpen
  \bibfield  {author} {\bibinfo {author} {\bibfnamefont {C.}~\bibnamefont
  {Husemann}}\ and\ \bibinfo {author} {\bibfnamefont {M.}~\bibnamefont
  {Salmhofer}},\ }\href {\doibase 10.1103/PhysRevB.79.195125} {\bibfield
  {journal} {\bibinfo  {journal} {Phys. Rev. B}\ }\textbf {\bibinfo {volume}
  {79}},\ \bibinfo {pages} {195125} (\bibinfo {year} {2009})}\BibitemShut
  {NoStop}%
\bibitem [{\citenamefont {Wang}\ \emph {et~al.}(2012)\citenamefont {Wang},
  \citenamefont {Xiang}, \citenamefont {Wang}, \citenamefont {Wang},
  \citenamefont {Yang},\ and\ \citenamefont {Lee}}]{Wang2012}%
  \BibitemOpen
  \bibfield  {author} {\bibinfo {author} {\bibfnamefont {W.-S.}\ \bibnamefont
  {Wang}}, \bibinfo {author} {\bibfnamefont {Y.-Y.}\ \bibnamefont {Xiang}},
  \bibinfo {author} {\bibfnamefont {Q.-H.}\ \bibnamefont {Wang}}, \bibinfo
  {author} {\bibfnamefont {F.}~\bibnamefont {Wang}}, \bibinfo {author}
  {\bibfnamefont {F.}~\bibnamefont {Yang}}, \ and\ \bibinfo {author}
  {\bibfnamefont {D.-H.}\ \bibnamefont {Lee}},\ }\href {\doibase
  10.1103/PhysRevB.85.035414} {\bibfield  {journal} {\bibinfo  {journal} {Phys.
  Rev. B}\ }\textbf {\bibinfo {volume} {85}},\ \bibinfo {pages} {035414}
  (\bibinfo {year} {2012})}\BibitemShut {NoStop}%
\bibitem [{\citenamefont {Zanchi}\ and\ \citenamefont
  {Schulz}(1998)}]{Zanchi1998}%
  \BibitemOpen
  \bibfield  {author} {\bibinfo {author} {\bibfnamefont {D.}~\bibnamefont
  {Zanchi}}\ and\ \bibinfo {author} {\bibfnamefont {H.~J.}\ \bibnamefont
  {Schulz}},\ }\href {http://stacks.iop.org/0295-5075/44/i=2/a=235} {\bibfield
  {journal} {\bibinfo  {journal} {EPL (Europhysics Letters)}\ }\textbf
  {\bibinfo {volume} {44}},\ \bibinfo {pages} {235} (\bibinfo {year}
  {1998})}\BibitemShut {NoStop}%
\bibitem [{\citenamefont {Honerkamp}\ \emph {et~al.}(2001)\citenamefont
  {Honerkamp}, \citenamefont {Salmhofer}, \citenamefont {Furukawa},\ and\
  \citenamefont {Rice}}]{Honerkamp2001}%
  \BibitemOpen
  \bibfield  {author} {\bibinfo {author} {\bibfnamefont {C.}~\bibnamefont
  {Honerkamp}}, \bibinfo {author} {\bibfnamefont {M.}~\bibnamefont
  {Salmhofer}}, \bibinfo {author} {\bibfnamefont {N.}~\bibnamefont {Furukawa}},
  \ and\ \bibinfo {author} {\bibfnamefont {T.~M.}\ \bibnamefont {Rice}},\
  }\href {\doibase 10.1103/PhysRevB.63.035109} {\bibfield  {journal} {\bibinfo
  {journal} {Phys. Rev. B}\ }\textbf {\bibinfo {volume} {63}},\ \bibinfo
  {pages} {035109} (\bibinfo {year} {2001})}\BibitemShut {NoStop}%
\bibitem [{\citenamefont {Halboth}\ and\ \citenamefont
  {Metzner}(2000)}]{Halboth2000}%
  \BibitemOpen
  \bibfield  {author} {\bibinfo {author} {\bibfnamefont {C.~J.}\ \bibnamefont
  {Halboth}}\ and\ \bibinfo {author} {\bibfnamefont {W.}~\bibnamefont
  {Metzner}},\ }\href {\doibase 10.1103/PhysRevB.61.7364} {\bibfield  {journal}
  {\bibinfo  {journal} {Phys. Rev. B}\ }\textbf {\bibinfo {volume} {61}},\
  \bibinfo {pages} {7364} (\bibinfo {year} {2000})}\BibitemShut {NoStop}%
\bibitem [{\citenamefont {{Lichtenstein}}\ \emph
  {et~al.}(2016{\natexlab{b}})\citenamefont {{Lichtenstein}}, \citenamefont
  {{Winkelmann}}, \citenamefont {{S{\'a}nchez de la Pe{\~n}a}}, \citenamefont
  {{Vidovi{\'c}}},\ and\ \citenamefont {{Di Napoli}}}]{Proceedings}%
  \BibitemOpen
  \bibfield  {author} {\bibinfo {author} {\bibfnamefont {J.}~\bibnamefont
  {{Lichtenstein}}}, \bibinfo {author} {\bibfnamefont {J.}~\bibnamefont
  {{Winkelmann}}}, \bibinfo {author} {\bibfnamefont {D.}~\bibnamefont
  {{S{\'a}nchez de la Pe{\~n}a}}}, \bibinfo {author} {\bibfnamefont
  {T.}~\bibnamefont {{Vidovi{\'c}}}}, \ and\ \bibinfo {author} {\bibfnamefont
  {E.}~\bibnamefont {{Di Napoli}}},\ }\href@noop {} {\bibfield  {journal}
  {\bibinfo  {journal} {ArXiv e-prints}\ } (\bibinfo {year}
  {2016}{\natexlab{b}})},\ \Eprint {http://arxiv.org/abs/1610.09991}
  {arXiv:1610.09991 [cs.CE]} \BibitemShut {NoStop}%
\bibitem [{\citenamefont {{Scherer}}\ \emph {et~al.}(2015)\citenamefont
  {{Scherer}}, \citenamefont {{Scherer}},\ and\ \citenamefont
  {{Honerkamp}}}]{Scherer2015}%
  \BibitemOpen
  \bibfield  {author} {\bibinfo {author} {\bibfnamefont {D.~D.}\ \bibnamefont
  {{Scherer}}}, \bibinfo {author} {\bibfnamefont {M.~M.}\ \bibnamefont
  {{Scherer}}}, \ and\ \bibinfo {author} {\bibfnamefont {C.}~\bibnamefont
  {{Honerkamp}}},\ }\href {\doibase 10.1103/PhysRevB.92.155137} {\bibfield
  {journal} {\bibinfo  {journal} {\prb}\ }\textbf {\bibinfo {volume} {92}},\
  \bibinfo {eid} {155137} (\bibinfo {year} {2015})}\BibitemShut {NoStop}%
\bibitem [{\citenamefont {Kiesel}\ \emph {et~al.}(2012)\citenamefont {Kiesel},
  \citenamefont {Platt}, \citenamefont {Hanke}, \citenamefont {Abanin},\ and\
  \citenamefont {Thomale}}]{Kiesel2012}%
  \BibitemOpen
  \bibfield  {author} {\bibinfo {author} {\bibfnamefont {M.~L.}\ \bibnamefont
  {Kiesel}}, \bibinfo {author} {\bibfnamefont {C.}~\bibnamefont {Platt}},
  \bibinfo {author} {\bibfnamefont {W.}~\bibnamefont {Hanke}}, \bibinfo
  {author} {\bibfnamefont {D.~A.}\ \bibnamefont {Abanin}}, \ and\ \bibinfo
  {author} {\bibfnamefont {R.}~\bibnamefont {Thomale}},\ }\href {\doibase
  10.1103/PhysRevB.86.020507} {\bibfield  {journal} {\bibinfo  {journal} {Phys.
  Rev. B}\ }\textbf {\bibinfo {volume} {86}},\ \bibinfo {pages} {020507}
  (\bibinfo {year} {2012})}\BibitemShut {NoStop}%
\bibitem [{\citenamefont {{Kiesel}}\ \emph {et~al.}(2013)\citenamefont
  {{Kiesel}}, \citenamefont {{Platt}},\ and\ \citenamefont
  {{Thomale}}}]{Kiesel2013}%
  \BibitemOpen
  \bibfield  {author} {\bibinfo {author} {\bibfnamefont {M.~L.}\ \bibnamefont
  {{Kiesel}}}, \bibinfo {author} {\bibfnamefont {C.}~\bibnamefont {{Platt}}}, \
  and\ \bibinfo {author} {\bibfnamefont {R.}~\bibnamefont {{Thomale}}},\ }\href
  {\doibase 10.1103/PhysRevLett.110.126405} {\bibfield  {journal} {\bibinfo
  {journal} {Physical Review Letters}\ }\textbf {\bibinfo {volume} {110}},\
  \bibinfo {eid} {126405} (\bibinfo {year} {2013})}\BibitemShut {NoStop}%
\bibitem [{\citenamefont {{Wang}}\ \emph {et~al.}(2013)\citenamefont {{Wang}},
  \citenamefont {{Li}}, \citenamefont {{Xiang}},\ and\ \citenamefont
  {{Wang}}}]{Wang2013}%
  \BibitemOpen
  \bibfield  {author} {\bibinfo {author} {\bibfnamefont {W.-S.}\ \bibnamefont
  {{Wang}}}, \bibinfo {author} {\bibfnamefont {Z.-Z.}\ \bibnamefont {{Li}}},
  \bibinfo {author} {\bibfnamefont {Y.-Y.}\ \bibnamefont {{Xiang}}}, \ and\
  \bibinfo {author} {\bibfnamefont {Q.-H.}\ \bibnamefont {{Wang}}},\ }\href
  {\doibase 10.1103/PhysRevB.87.115135} {\bibfield  {journal} {\bibinfo
  {journal} {\prb}\ }\textbf {\bibinfo {volume} {87}},\ \bibinfo {eid} {115135}
  (\bibinfo {year} {2013})}\BibitemShut {NoStop}%
\bibitem [{\citenamefont {Raghu}\ \emph {et~al.}(2008)\citenamefont {Raghu},
  \citenamefont {Qi}, \citenamefont {Honerkamp},\ and\ \citenamefont
  {Zhang}}]{Raghu2008}%
  \BibitemOpen
  \bibfield  {author} {\bibinfo {author} {\bibfnamefont {S.}~\bibnamefont
  {Raghu}}, \bibinfo {author} {\bibfnamefont {X.-L.}\ \bibnamefont {Qi}},
  \bibinfo {author} {\bibfnamefont {C.}~\bibnamefont {Honerkamp}}, \ and\
  \bibinfo {author} {\bibfnamefont {S.-C.}\ \bibnamefont {Zhang}},\ }\href
  {\doibase 10.1103/PhysRevLett.100.156401} {\bibfield  {journal} {\bibinfo
  {journal} {Phys. Rev. Lett.}\ }\textbf {\bibinfo {volume} {100}},\ \bibinfo
  {pages} {156401} (\bibinfo {year} {2008})}\BibitemShut {NoStop}%
\bibitem [{\citenamefont {Wehling}\ \emph {et~al.}(2011)\citenamefont
  {Wehling}, \citenamefont {\ifmmode \mbox{\c{S}}\else \c{S}\fi{}a\ifmmode
  \mbox{\c{s}}\else \c{s}\fi{}\ifmmode \imath \else \i
  \fi{}o\ifmmode~\breve{g}\else \u{g}\fi{}lu}, \citenamefont {Friedrich},
  \citenamefont {Lichtenstein}, \citenamefont {Katsnelson},\ and\ \citenamefont
  {Bl\"ugel}}]{Wehling2011}%
  \BibitemOpen
  \bibfield  {author} {\bibinfo {author} {\bibfnamefont {T.~O.}\ \bibnamefont
  {Wehling}}, \bibinfo {author} {\bibfnamefont {E.}~\bibnamefont {\ifmmode
  \mbox{\c{S}}\else \c{S}\fi{}a\ifmmode \mbox{\c{s}}\else \c{s}\fi{}\ifmmode
  \imath \else \i \fi{}o\ifmmode~\breve{g}\else \u{g}\fi{}lu}}, \bibinfo
  {author} {\bibfnamefont {C.}~\bibnamefont {Friedrich}}, \bibinfo {author}
  {\bibfnamefont {A.~I.}\ \bibnamefont {Lichtenstein}}, \bibinfo {author}
  {\bibfnamefont {M.~I.}\ \bibnamefont {Katsnelson}}, \ and\ \bibinfo {author}
  {\bibfnamefont {S.}~\bibnamefont {Bl\"ugel}},\ }\href {\doibase
  10.1103/PhysRevLett.106.236805} {\bibfield  {journal} {\bibinfo  {journal}
  {Phys. Rev. Lett.}\ }\textbf {\bibinfo {volume} {106}},\ \bibinfo {pages}
  {236805} (\bibinfo {year} {2011})}\BibitemShut {NoStop}%
\bibitem [{\citenamefont {Wetterich}(1993)}]{Wetterich1993}%
  \BibitemOpen
  \bibfield  {author} {\bibinfo {author} {\bibfnamefont {C.}~\bibnamefont
  {Wetterich}},\ }\href {\doibase
  http://dx.doi.org/10.1016/0370-2693(93)90726-X} {\bibfield  {journal}
  {\bibinfo  {journal} {Physics Letters B}\ }\textbf {\bibinfo {volume}
  {301}},\ \bibinfo {pages} {90 } (\bibinfo {year} {1993})}\BibitemShut
  {NoStop}%
\bibitem [{\citenamefont {Metzner}\ \emph {et~al.}(2012)\citenamefont
  {Metzner}, \citenamefont {Salmhofer}, \citenamefont {Honerkamp},
  \citenamefont {Meden},\ and\ \citenamefont {Sch\"onhammer}}]{Metzner2012}%
  \BibitemOpen
  \bibfield  {author} {\bibinfo {author} {\bibfnamefont {W.}~\bibnamefont
  {Metzner}}, \bibinfo {author} {\bibfnamefont {M.}~\bibnamefont {Salmhofer}},
  \bibinfo {author} {\bibfnamefont {C.}~\bibnamefont {Honerkamp}}, \bibinfo
  {author} {\bibfnamefont {V.}~\bibnamefont {Meden}}, \ and\ \bibinfo {author}
  {\bibfnamefont {K.}~\bibnamefont {Sch\"onhammer}},\ }\href {\doibase
  10.1103/RevModPhys.84.299} {\bibfield  {journal} {\bibinfo  {journal} {Rev.
  Mod. Phys.}\ }\textbf {\bibinfo {volume} {84}},\ \bibinfo {pages} {299}
  (\bibinfo {year} {2012})}\BibitemShut {NoStop}%
\bibitem [{\citenamefont {{Platt}}\ \emph {et~al.}(2013)\citenamefont
  {{Platt}}, \citenamefont {{Hanke}},\ and\ \citenamefont
  {{Thomale}}}]{Thomale2013}%
  \BibitemOpen
  \bibfield  {author} {\bibinfo {author} {\bibfnamefont {C.}~\bibnamefont
  {{Platt}}}, \bibinfo {author} {\bibfnamefont {W.}~\bibnamefont {{Hanke}}}, \
  and\ \bibinfo {author} {\bibfnamefont {R.}~\bibnamefont {{Thomale}}},\ }\href
  {\doibase 10.1080/00018732.2013.862020} {\bibfield  {journal} {\bibinfo
  {journal} {Advances in Physics}\ }\textbf {\bibinfo {volume} {62}},\ \bibinfo
  {pages} {453} (\bibinfo {year} {2013})}\BibitemShut {NoStop}%
\bibitem [{\citenamefont {Salmhofer}\ and\ \citenamefont
  {Honerkamp}(2001)}]{Salmhofer2001}%
  \BibitemOpen
  \bibfield  {author} {\bibinfo {author} {\bibfnamefont {M.}~\bibnamefont
  {Salmhofer}}\ and\ \bibinfo {author} {\bibfnamefont {C.}~\bibnamefont
  {Honerkamp}},\ }\href {\doibase 10.1143/PTP.105.1} {\bibfield  {journal}
  {\bibinfo  {journal} {Progress of Theoretical Physics}\ }\textbf {\bibinfo
  {volume} {105}},\ \bibinfo {pages} {1} (\bibinfo {year} {2001})}\BibitemShut
  {NoStop}%
\bibitem [{\citenamefont {Husemann}\ \emph {et~al.}(2012)\citenamefont
  {Husemann}, \citenamefont {Giering},\ and\ \citenamefont
  {Salmhofer}}]{Husemann2012}%
  \BibitemOpen
  \bibfield  {author} {\bibinfo {author} {\bibfnamefont {C.}~\bibnamefont
  {Husemann}}, \bibinfo {author} {\bibfnamefont {K.-U.}\ \bibnamefont
  {Giering}}, \ and\ \bibinfo {author} {\bibfnamefont {M.}~\bibnamefont
  {Salmhofer}},\ }\href {\doibase 10.1103/PhysRevB.85.075121} {\bibfield
  {journal} {\bibinfo  {journal} {Phys. Rev. B}\ }\textbf {\bibinfo {volume}
  {85}},\ \bibinfo {pages} {075121} (\bibinfo {year} {2012})}\BibitemShut
  {NoStop}%
\bibitem [{\citenamefont {Eberlein}\ and\ \citenamefont
  {Metzner}(2010)}]{Eberlein2010}%
  \BibitemOpen
  \bibfield  {author} {\bibinfo {author} {\bibfnamefont {A.}~\bibnamefont
  {Eberlein}}\ and\ \bibinfo {author} {\bibfnamefont {W.}~\bibnamefont
  {Metzner}},\ }\href {\doibase 10.1143/PTP.124.471} {\bibfield  {journal}
  {\bibinfo  {journal} {Progress of Theoretical Physics}\ }\textbf {\bibinfo
  {volume} {124}},\ \bibinfo {pages} {471} (\bibinfo {year}
  {2010})}\BibitemShut {NoStop}%
\bibitem [{\citenamefont {{Eberlein}}\ and\ \citenamefont
  {{Metzner}}(2013)}]{Eberlein2013}%
  \BibitemOpen
  \bibfield  {author} {\bibinfo {author} {\bibfnamefont {A.}~\bibnamefont
  {{Eberlein}}}\ and\ \bibinfo {author} {\bibfnamefont {W.}~\bibnamefont
  {{Metzner}}},\ }\href {\doibase 10.1103/PhysRevB.87.174523} {\bibfield
  {journal} {\bibinfo  {journal} {\prb}\ }\textbf {\bibinfo {volume} {87}},\
  \bibinfo {eid} {174523} (\bibinfo {year} {2013})}\BibitemShut {NoStop}%
\bibitem [{\citenamefont {Maier}\ \emph {et~al.}(2013)\citenamefont {Maier},
  \citenamefont {Ortloff},\ and\ \citenamefont {Honerkamp}}]{Maier2013}%
  \BibitemOpen
  \bibfield  {author} {\bibinfo {author} {\bibfnamefont {S.~A.}\ \bibnamefont
  {Maier}}, \bibinfo {author} {\bibfnamefont {J.}~\bibnamefont {Ortloff}}, \
  and\ \bibinfo {author} {\bibfnamefont {C.}~\bibnamefont {Honerkamp}},\ }\href
  {\doibase 10.1103/PhysRevB.88.235112} {\bibfield  {journal} {\bibinfo
  {journal} {Phys. Rev. B}\ }\textbf {\bibinfo {volume} {88}},\ \bibinfo
  {pages} {235112} (\bibinfo {year} {2013})}\BibitemShut {NoStop}%
\bibitem [{\citenamefont {Xiang}\ \emph {et~al.}(2013)\citenamefont {Xiang},
  \citenamefont {Yang}, \citenamefont {Wang}, \citenamefont {Li},\ and\
  \citenamefont {Wang}}]{Xiang2013}%
  \BibitemOpen
  \bibfield  {author} {\bibinfo {author} {\bibfnamefont {Y.-Y.}\ \bibnamefont
  {Xiang}}, \bibinfo {author} {\bibfnamefont {Y.}~\bibnamefont {Yang}},
  \bibinfo {author} {\bibfnamefont {W.-S.}\ \bibnamefont {Wang}}, \bibinfo
  {author} {\bibfnamefont {Z.-Z.}\ \bibnamefont {Li}}, \ and\ \bibinfo {author}
  {\bibfnamefont {Q.-H.}\ \bibnamefont {Wang}},\ }\href {\doibase
  10.1103/PhysRevB.88.104516} {\bibfield  {journal} {\bibinfo  {journal} {Phys.
  Rev. B}\ }\textbf {\bibinfo {volume} {88}},\ \bibinfo {pages} {104516}
  (\bibinfo {year} {2013})}\BibitemShut {NoStop}%
\bibitem [{\citenamefont {{Eberlein}}(2014)}]{Eberlein2014}%
  \BibitemOpen
  \bibfield  {author} {\bibinfo {author} {\bibfnamefont {A.}~\bibnamefont
  {{Eberlein}}},\ }\href {\doibase 10.1103/PhysRevB.90.115125} {\bibfield
  {journal} {\bibinfo  {journal} {\prb}\ }\textbf {\bibinfo {volume} {90}},\
  \bibinfo {eid} {115125} (\bibinfo {year} {2014})}\BibitemShut {NoStop}%
\bibitem [{\citenamefont {Volpez}\ \emph {et~al.}(2016)\citenamefont {Volpez},
  \citenamefont {Scherer},\ and\ \citenamefont {Scherer}}]{Yanick2016}%
  \BibitemOpen
  \bibfield  {author} {\bibinfo {author} {\bibfnamefont {Y.}~\bibnamefont
  {Volpez}}, \bibinfo {author} {\bibfnamefont {D.~D.}\ \bibnamefont {Scherer}},
  \ and\ \bibinfo {author} {\bibfnamefont {M.~M.}\ \bibnamefont {Scherer}},\
  }\href {\doibase 10.1103/PhysRevB.94.165107} {\bibfield  {journal} {\bibinfo
  {journal} {Phys. Rev. B}\ }\textbf {\bibinfo {volume} {94}},\ \bibinfo
  {pages} {165107} (\bibinfo {year} {2016})}\BibitemShut {NoStop}%
\bibitem [{\citenamefont {Scherer}\ \emph
  {et~al.}(2012{\natexlab{a}})\citenamefont {Scherer}, \citenamefont
  {Uebelacker},\ and\ \citenamefont {Honerkamp}}]{Scherer2012}%
  \BibitemOpen
  \bibfield  {author} {\bibinfo {author} {\bibfnamefont {M.~M.}\ \bibnamefont
  {Scherer}}, \bibinfo {author} {\bibfnamefont {S.}~\bibnamefont {Uebelacker}},
  \ and\ \bibinfo {author} {\bibfnamefont {C.}~\bibnamefont {Honerkamp}},\
  }\href {\doibase 10.1103/PhysRevB.85.235408} {\bibfield  {journal} {\bibinfo
  {journal} {Phys. Rev. B}\ }\textbf {\bibinfo {volume} {85}},\ \bibinfo
  {pages} {235408} (\bibinfo {year} {2012}{\natexlab{a}})}\BibitemShut
  {NoStop}%
\bibitem [{\citenamefont {Scherer}\ \emph
  {et~al.}(2012{\natexlab{b}})\citenamefont {Scherer}, \citenamefont
  {Uebelacker}, \citenamefont {Scherer},\ and\ \citenamefont
  {Honerkamp}}]{Scherer2012a}%
  \BibitemOpen
  \bibfield  {author} {\bibinfo {author} {\bibfnamefont {M.~M.}\ \bibnamefont
  {Scherer}}, \bibinfo {author} {\bibfnamefont {S.}~\bibnamefont {Uebelacker}},
  \bibinfo {author} {\bibfnamefont {D.~D.}\ \bibnamefont {Scherer}}, \ and\
  \bibinfo {author} {\bibfnamefont {C.}~\bibnamefont {Honerkamp}},\ }\href
  {\doibase 10.1103/PhysRevB.86.155415} {\bibfield  {journal} {\bibinfo
  {journal} {Phys. Rev. B}\ }\textbf {\bibinfo {volume} {86}},\ \bibinfo
  {pages} {155415} (\bibinfo {year} {2012}{\natexlab{b}})}\BibitemShut
  {NoStop}%
\bibitem [{\citenamefont {{Jia}}\ \emph {et~al.}(2013)\citenamefont {{Jia}},
  \citenamefont {{Guo}}, \citenamefont {{Chen}}, \citenamefont {{Shen}},\ and\
  \citenamefont {{Feng}}}]{Jia2013}%
  \BibitemOpen
  \bibfield  {author} {\bibinfo {author} {\bibfnamefont {Y.}~\bibnamefont
  {{Jia}}}, \bibinfo {author} {\bibfnamefont {H.}~\bibnamefont {{Guo}}},
  \bibinfo {author} {\bibfnamefont {Z.}~\bibnamefont {{Chen}}}, \bibinfo
  {author} {\bibfnamefont {S.-Q.}\ \bibnamefont {{Shen}}}, \ and\ \bibinfo
  {author} {\bibfnamefont {S.}~\bibnamefont {{Feng}}},\ }\href {\doibase
  10.1103/PhysRevB.88.075101} {\bibfield  {journal} {\bibinfo  {journal}
  {\prb}\ }\textbf {\bibinfo {volume} {88}},\ \bibinfo {eid} {075101} (\bibinfo
  {year} {2013})}\BibitemShut {NoStop}%
\bibitem [{\citenamefont {Daghofer}\ and\ \citenamefont
  {Hohenadler}(2014)}]{Daghofer2014}%
  \BibitemOpen
  \bibfield  {author} {\bibinfo {author} {\bibfnamefont {M.}~\bibnamefont
  {Daghofer}}\ and\ \bibinfo {author} {\bibfnamefont {M.}~\bibnamefont
  {Hohenadler}},\ }\href {\doibase 10.1103/PhysRevB.89.035103} {\bibfield
  {journal} {\bibinfo  {journal} {Phys. Rev. B}\ }\textbf {\bibinfo {volume}
  {89}},\ \bibinfo {pages} {035103} (\bibinfo {year} {2014})}\BibitemShut
  {NoStop}%
\bibitem [{\citenamefont {{Capponi}}\ and\ \citenamefont
  {{L{\"a}uchli}}(2015)}]{Capponi2015}%
  \BibitemOpen
  \bibfield  {author} {\bibinfo {author} {\bibfnamefont {S.}~\bibnamefont
  {{Capponi}}}\ and\ \bibinfo {author} {\bibfnamefont {A.~M.}\ \bibnamefont
  {{L{\"a}uchli}}},\ }\href {\doibase 10.1103/PhysRevB.92.085146} {\bibfield
  {journal} {\bibinfo  {journal} {\prb}\ }\textbf {\bibinfo {volume} {92}},\
  \bibinfo {eid} {085146} (\bibinfo {year} {2015})}\BibitemShut {NoStop}%
\bibitem [{\citenamefont {{Motruk}}\ \emph {et~al.}(2015)\citenamefont
  {{Motruk}}, \citenamefont {{Grushin}}, \citenamefont {{de Juan}},\ and\
  \citenamefont {{Pollmann}}}]{Motruk2015}%
  \BibitemOpen
  \bibfield  {author} {\bibinfo {author} {\bibfnamefont {J.}~\bibnamefont
  {{Motruk}}}, \bibinfo {author} {\bibfnamefont {A.~G.}\ \bibnamefont
  {{Grushin}}}, \bibinfo {author} {\bibfnamefont {F.}~\bibnamefont {{de
  Juan}}}, \ and\ \bibinfo {author} {\bibfnamefont {F.}~\bibnamefont
  {{Pollmann}}},\ }\href {\doibase 10.1103/PhysRevB.92.085147} {\bibfield
  {journal} {\bibinfo  {journal} {\prb}\ }\textbf {\bibinfo {volume} {92}},\
  \bibinfo {eid} {085147} (\bibinfo {year} {2015})}\BibitemShut {NoStop}%
\bibitem [{\citenamefont {S\'anchez de~la Pe\~na}\ \emph
  {et~al.}(2014)\citenamefont {S\'anchez de~la Pe\~na}, \citenamefont
  {Scherer},\ and\ \citenamefont {Honerkamp}}]{Pena2014}%
  \BibitemOpen
  \bibfield  {author} {\bibinfo {author} {\bibfnamefont {D.}~\bibnamefont
  {S\'anchez de~la Pe\~na}}, \bibinfo {author} {\bibfnamefont {M.~M.}\
  \bibnamefont {Scherer}}, \ and\ \bibinfo {author} {\bibfnamefont
  {C.}~\bibnamefont {Honerkamp}},\ }\href {\doibase 10.1002/andp.201400088}
  {\bibfield  {journal} {\bibinfo  {journal} {Annalen der Physik}\ }\textbf
  {\bibinfo {volume} {526}},\ \bibinfo {pages} {366} (\bibinfo {year}
  {2014})}\BibitemShut {NoStop}%
\bibitem [{\citenamefont {{Maier}}\ \emph {et~al.}(2014)\citenamefont
  {{Maier}}, \citenamefont {{Eberlein}},\ and\ \citenamefont
  {{Honerkamp}}}]{Maier2014}%
  \BibitemOpen
  \bibfield  {author} {\bibinfo {author} {\bibfnamefont {S.~A.}\ \bibnamefont
  {{Maier}}}, \bibinfo {author} {\bibfnamefont {A.}~\bibnamefont {{Eberlein}}},
  \ and\ \bibinfo {author} {\bibfnamefont {C.}~\bibnamefont {{Honerkamp}}},\
  }\href {\doibase 10.1103/PhysRevB.90.035140} {\bibfield  {journal} {\bibinfo
  {journal} {\prb}\ }\textbf {\bibinfo {volume} {90}},\ \bibinfo {eid} {035140}
  (\bibinfo {year} {2014})}\BibitemShut {NoStop}%
\bibitem [{\citenamefont {{Maier}}\ \emph {et~al.}(2013)\citenamefont
  {{Maier}}, \citenamefont {{Honerkamp}},\ and\ \citenamefont
  {{Wang}}}]{Maier2013a}%
  \BibitemOpen
  \bibfield  {author} {\bibinfo {author} {\bibfnamefont {S.~A.}\ \bibnamefont
  {{Maier}}}, \bibinfo {author} {\bibfnamefont {C.}~\bibnamefont
  {{Honerkamp}}}, \ and\ \bibinfo {author} {\bibfnamefont {Q.-H.}\ \bibnamefont
  {{Wang}}},\ }\href@noop {} {\bibfield  {journal} {\bibinfo  {journal}
  {Symmetry}\ }\textbf {\bibinfo {volume} {4}},\ \bibinfo {pages} {313}
  (\bibinfo {year} {2013})}\BibitemShut {NoStop}%
\bibitem [{\citenamefont {Ahnert}\ and\ \citenamefont
  {Mulansky}(2011)}]{Ahnert2011}%
  \BibitemOpen
  \bibfield  {author} {\bibinfo {author} {\bibfnamefont {K.}~\bibnamefont
  {Ahnert}}\ and\ \bibinfo {author} {\bibfnamefont {M.}~\bibnamefont
  {Mulansky}},\ }\href {\doibase http://dx.doi.org/10.1063/1.3637934}
  {\bibfield  {journal} {\bibinfo  {journal} {AIP Conference Proceedings}\
  }\textbf {\bibinfo {volume} {1389}},\ \bibinfo {pages} {1586} (\bibinfo
  {year} {2011})}\BibitemShut {NoStop}%
\bibitem [{\citenamefont {Berntsen}\ \emph {et~al.}(1991)\citenamefont
  {Berntsen}, \citenamefont {Espelid},\ and\ \citenamefont
  {Genz}}]{Berntsen1991}%
  \BibitemOpen
  \bibfield  {author} {\bibinfo {author} {\bibfnamefont {J.}~\bibnamefont
  {Berntsen}}, \bibinfo {author} {\bibfnamefont {T.~O.}\ \bibnamefont
  {Espelid}}, \ and\ \bibinfo {author} {\bibfnamefont {A.}~\bibnamefont
  {Genz}},\ }\href {\doibase 10.1145/210232.210234} {\bibfield  {journal}
  {\bibinfo  {journal} {ACM Trans. Math. Softw.}\ }\textbf {\bibinfo {volume}
  {17}},\ \bibinfo {pages} {452} (\bibinfo {year} {1991})}\BibitemShut
  {NoStop}%
\bibitem [{\citenamefont {Luehrs}\ \emph {et~al.}(2016)\citenamefont {Luehrs},
  \citenamefont {Rohe}, \citenamefont {Schnurpfeil}, \citenamefont {Thust},\
  and\ \citenamefont {Frings}}]{Luhrs2016}%
  \BibitemOpen
  \bibfield  {author} {\bibinfo {author} {\bibfnamefont {S.}~\bibnamefont
  {Luehrs}}, \bibinfo {author} {\bibfnamefont {D.}~\bibnamefont {Rohe}},
  \bibinfo {author} {\bibfnamefont {A.}~\bibnamefont {Schnurpfeil}}, \bibinfo
  {author} {\bibfnamefont {K.}~\bibnamefont {Thust}}, \ and\ \bibinfo {author}
  {\bibfnamefont {W.}~\bibnamefont {Frings}},\ }\href {\doibase
  10.3233/978-1-61499-621-7-431} {\bibfield  {journal} {\bibinfo  {journal}
  {Advances in Parallel Computing}\ }\textbf {\bibinfo {volume} {27}},\
  \bibinfo {pages} {431–438} (\bibinfo {year} {2016})}\BibitemShut {NoStop}%
\end{thebibliography}%

\end{document}